\documentclass[useAMS,usenatbib]{mnras}
\usepackage{newtxtext,newtxmath}
\usepackage[T1]{fontenc}
\usepackage{ae,aecompl}
\usepackage{graphicx}	           % Including figure files
\usepackage{amsmath}	% Advanced maths commands
\usepackage{amssymb}	% Extra maths symbols
\usepackage{natbib}
\usepackage[para,online,flushleft]{threeparttable}
\usepackage{booktabs}
\usepackage{color}
\definecolor{modif}{rgb}{0,0.7,0}
\definecolor{comment}{rgb}{1,0,0}
\newcommand{\kms}{km\,s$^{-1}$}
\newcommand{\xic}{$\xi^1$\,CMa}
\newcommand{\oc}{O\,$-$\,C}
\newcommand{\GG}[1]{}

\definecolor{darkblue}{rgb}{0.0, 0.0, 0.60}
\newcommand{\msun}{M$_{\odot}$}
\newcommand{\bz}{\ensuremath{\langle B_z\rangle}}

\def\gtrsim{\mathrel{\hbox{\rlap{\hbox{\lower4pt\hbox{$\sim$}}}\hbox{$>$}}}}
\def\ltsim{\mathrel{\hbox{\rlap{\hbox{\lower4pt\hbox{$\sim$}}}\hbox{$<$}}}}

\title[{\xic}]{Evolving pulsation of the slowly rotating magnetic $\beta$ Cep star \xic}

\author[G.A. Wade et al.]{G.A. Wade\thanks{E-mail: wade-g@rmc.ca}$^1$, A. Pigulski$^2$, S. Begy$^1$, M. Shultz$^3$, G. Handler$^4$, J. Sikora$^5$, 
\newauthor{H. Neilson$^6$, H. Cugier$^2$, C. Erba$^3$, A.F.J. Moffat$^7$, B. Pablo$^8$, A. Popowicz$^9$,}
\newauthor{W. Weiss$^{10}$, K. Zwintz$^{11}$}
\\
$^{1}$Dept. of Physics \& Space Science, Royal Military College of Canada, PO Box 17000 Station Forces, Kingston, ON, Canada K7K 0C6 \\
$^2$Instytut Astronomiczny, Uniwersytet Wroc{\l}awski, Kopernika 11, 51-622 Wroc{\l}aw, Poland\\
$^3$Dept. of Physics and Astronomy, University of Delaware, 217 Sharp Lab, Newark, DE 19716, USA\\
$^4$Nicolaus Copernicus Astronomical Center, Bartycka 18, 00--716 Warszawa, Poland\\
$^5$Dept. of Physics, Engineering Physics and Astronomy, Queen's University, 99 University Avenue, Kingston, ON K7L 3N6, Canada\\
$^6$Dept. of Astronomy \& Astrophysics, University of Toronto, 50 St. George Street, Toronto, ON M5S 3H4, Canada\\
$^7$D\'ept. de physique and Centre de Recherche en Astrophysique du Qu\'ebec (CRAQ), Universit\'e de Montr\'eal, C.P. 6128, \\\ \ Succ. Centre-Ville, Montr\'eal, QC H3C 3J7, Canada\\
$^8$American Association of Variable Star Observers, 49 Bay State Road, Cambridge, MA 02138, USA\\
$^9$Silesian University of Technology, Institute of Automatic Control, Akademicka 16, Gliwice, Poland\\
$^{10}$Institut f\"ur Astrophysik, Universit\"at Wien, T\"urkenschanzstrasse 17, A-1180 Wien, Austria\\
$^{11}$Institut f\"ur Astro- und Teilchenphysik, Universit\"at Innsbruck, Technikerstrasse 25/8, A-6020 Innsbruck, Austria
}

\date{Accepted . Received ; in original form }

\pubyear{2019}

\begin{document}
\label{firstpage}
\pagerange{\pageref{firstpage}--\pageref{lastpage}}
\maketitle

\begin{abstract}
Recent BRITE-Constellation space photometry of the slowly rotating, magnetic $\beta$~Cep pulsator \xic\ permits a new analysis of its pulsation properties. Analysis of the two-colour BRITE data reveals the well-known single pulsation period of $0.209$~d, along with its first and second harmonics. A similar analysis of SMEI and TESS observations yields compatible results, with the higher precision TESS observations also revealing several low-amplitude modes with frequencies below 5~d$^{-1}$; some of these are likely $g$ modes. The phase lag between photometric and radial velocity maxima - equal to 0.334 cycles - is significantly larger than the typical value of $1/4$ observed in other large-amplitude $\beta$~Cep stars. The phase lag, as well as the strong dependence of phase of maximum light on wavelength, can be reconciled with seismic models only if the dominant mode is the fundamental radial mode. We employ all published photometric and radial velocity measurements, spanning over a century, to evaluate the stability of the pulsation period. The $O-C$ diagram exhibits a clear parabolic shape consistent with a mean rate of period change {$\dot P=0.34\pm 0.02$~s/cen}. The residuals from the best-fit parabola exhibit scatter that is substantially larger than the uncertainties. In particular, dense sampling obtained during the past $\sim$20 years suggests more complex and rapid period variations. Those data cannot be coherently phased with the mean rate of period change, and instead require $\dot P\sim0.9$~s/cen. We examine the potential contributions of binarity, stellar evolution, and stellar rotation and magnetism to understand the apparent period evolution. %We find no evidence for any other frequencies with a best upper limit of 0.17~mmag. 
\end{abstract}


%\author{Sean B. Begy}[RMC]
%\author{Gregg A. Wade}[RMC]
%\author{Gerald Handler}[CAMK]
%\author{Andrzej Pigulski}[Wroclaw]
%\author{James Sikora}[RMC]
%\author{Matt Shultz}[Uppsala]
%\author{the BRITE team}
%\affil[RMC]{Department of Physics, Royal Military College of Canada, Kingston, Ontario K7K 7B4, Canada}
%\affil[CAMK]{Nicolaus Copernicus Astronomical Center, Bartycka 18, 00--716 Warszawa, Poland}
%\affil[Uppsala]{Department of Physics and Astronomy, Uppsala University, Box 516, Uppsala 75120}
%\affil[Wroclaw]{Instytut Astronomiczny, Uniwersytet Wroclawski, Kopernika 11, PL-51-622 Wroclaw, Poland}
%\title{Evolving pulsation of the slowly rotating magnetic $\beta$ Cep star {\xic}}

\section{Introduction}

{\xic} (HR\,2387 = HD\,46328 = HIP\,31125 = MCW 441 = ADS\,5176A) is a bright ($V=4.3$\,mag), apparently single, B0.5 subgiant located near the middle of its main sequence evolution \citep[e.g.][]{2005A&A...433..659N,2017MNRAS.471.2286S}. It has long been known to exhibit large-amplitude $\beta$~Cep radial pulsations with a period of roughly $P=0.20958$~d \citep[4.77 d$^{-1}$; e.g.][]{1953PASP...65..193M,1954PASP...66..200W,1992AAS...96..207H, 2006CoAst.147..109S}.

\citet{2009RMxAC..36..319H} reported the star to be magnetic. \citet{2017MNRAS.471.2286S} performed a detailed determination of the star's physical parameters, finding $T_{\rm eff}=27\pm 1$~kK, $\log g=3.78\pm 0.07$, and an age of $11.1\pm 0.7$~Myr. At an inferred mass of $14.2\pm 0.4$~M$_\odot$, this implies that {\xic} has completed three-quarters of its main sequence evolution. 

Analysis of high resolution spectropolarimetry obtained between 2000\,--\,2017 led \citet{2017MNRAS.471.2286S} to conclude that the star's rotation period is remarkably long, over 30 years. In particular, \citet{2017MNRAS.471.2286S} \citep[see as well][]{2018MNRAS.478L..39S} demonstrated that previous claims of much shorter rotation periods were unable to explain the magnetic observations. \cite{2018MNRAS.478L..39S} discovered the presence of unexpected crossover signatures in Stokes $V$ profiles of \xic\ obtained near the phase of null longitudinal field. They demonstrated that the combination of radial pulsation and departures from a dipole magnetic field geometry could explain the presence of this novel and unexpected ``radial crossover'' effect.  
 
\citet{2017MNRAS.471.2286S} also examined the behaviour of the radial-velocity (RV) pulsations of the star over a span of 17~yr. They demonstrated that a constant pulsation period was unable to phase those data coherently, and consequently inferred that the period was increasing at a rate of 0.96 s/cen. This result is qualitatively consistent with earlier reports of period instability of \xic. \citet{1999NewAR..43..455J}, in his summary of period evolution of $\beta$~Cep stars, cites a rate of period change of $0.37\pm 0.05$~s/cen reported {by \cite{1992PhDT-Pigulski}}. \citet{2015A&A...584A..58N} used those results to test the influence of rotation and convective core overshoot on models of massive star evolution, finding that the measured rate of period change of \xic\ was in good agreement with that predicted by models under the constraints applied by the physical parameters of \citet{2016PhDT.......390S,2017MNRAS.471.2286S}. %\comment{[API asks "Is "qualitatively" appropriate if the period change rates differ by a factor of 3? GAW replies "qualitatively" in this context means that the the change has the correct sign and behaviour.]}

Real or apparent pulsation period evolution can be the consequence of a number of phenomena, including binarity and stellar evolution.  For example, \cite{1984PASP...96..657O, 2012A&A...544A..28O} reported that the $\beta$~Cep star BW~Vul exhibits complex period variability that \citet{2012A&A...544A..28O} concluded is best understood as a piecewise linear ephemeris, corresponding to a constant period interrupted every few decades by an abrupt period change. A number of studies have considered the role of stellar evolution and binarity in understanding pulsation period changes \citep{2016JAVSO..44..179N}. \cite{1919Obs....42..338E} conducted the first test by considering the rate of period change for the prototype Cepheid, $\delta$~Cephei, and showed that the rate of period change was inconsistent with energy generation from gravitational contraction. More recently, numerous works have used period change measurements to test evolution of Cepheids such as Polaris, $\delta$~Cep, and l~Car \citep{2012ApJ...745L..32N,2016JAVSO..44..179N,2014A&A...563A..48N,2015MNRAS.449.1011F,2018A&A...611L...7A}. \cite{2015ApJ...804..144A} considered the period change of $\delta$~Cep as potential evidence for an undetected close companion.  Period change due to evolutionary effects has not been examined in detail for RR~Lyrae stars; however \cite{1994ApJ...423..380K} and \cite{2011AJ....141...15K} made predictions of period change from stellar evolution models and found some consistency with observations.  This has been confirmed by \cite{2007A&A...476..307L} and  \cite{2013JAVSO..41...75P}.  The light-time effect due to binary companions appears to be one of the origins of apparent period changes for the $\beta$~Cephei stars \citep{1992A&A...253..178P,1992A&A...261..203P,1993A&A...274..269P}.

In this paper we revisit (a) the pulsation frequency spectrum and (b) the evolution of the fundamental pulsation period of \xic. We report new two-colour BRITE photometry of the star which we analyse in tandem with SMEI photometry to search for evidence of additional pulsation frequencies. We then revisit the period evolution reported by \citet{1999NewAR..43..455J} and \citet{2017MNRAS.471.2286S} using all published photometric and radial velocity (RV) measurements of \xic, spanning over 100 years.

%\modif{State that this is the fundamental radial mode.}

%\section{Variability}
%4 {\xic} = HR\,2387 = HD\,46328 ($V=$ 4.8~mag) was found to be variable in radial velocities by \cite{1907ApJ....25R..59F}, which was later confirmed by \cite{1921PDO.....5...45H}. Given the rapid changes of RVs, \cite{1921PDO.....5...45H} included it into the $\beta$~CMa ($\beta$~Cep) group of variable stars supposed to be binaries at that time. The variability period, equal to about 0.2096~d, was first derived from RVs by \cite{1953PASP...65..193M} and confirmed in photometry by \cite{1954PASP...66..200W}.

\begin{table*}
\begin{flushleft}
\caption[]{\label{tab:RV_table} New (2018-2019) radial velocity measurements of $\xi^1$CMa. These data are described in Sect.~\ref{specpol}.}
\end{flushleft}
\begin{center}
\begin{tabular}{c c|c c|c c}
\hline
%\hline
HJD-2458000 & RV (km s$^{-1}$) & HJD-2458000 & RV (km s$^{-1}$) & HJD-2458000 & RV (km s$^{-1}$) \\ %\midrule
\hline
148.75321   & $11.8 \pm 0.7$        & 557.77240   & $39.9 \pm 1.3$        & 559.79702   & $14.5 \pm 0.6$        \\
148.75429   & $11.6 \pm 0.7$        & 557.77349   & $40.1 \pm 1.3$        & 559.81117   & $20.4 \pm 0.8$        \\
148.75537   & $10.0 \pm 0.7$        & 557.77459   & $40.0 \pm 1.3$        & 559.81230   & $20.9 \pm 0.8$        \\
148.75645   & $10.2 \pm 0.7$        & 557.77569   & $40.0 \pm 1.3$        & 559.81342   & $21.5 \pm 0.8$        \\
148.97144   & $9.5 \pm 0.7$         & 557.83743   & $16.4 \pm 0.7$        & 559.81455   & $21.9 \pm 0.8$        \\
148.97252   & $9.3 \pm 0.7$         & 557.83856   & $15.9 \pm 0.7$        & 559.82356   & $26.1 \pm 0.9$        \\
148.97360   & $9.8 \pm 0.7$         & 557.83969   & $15.4 \pm 0.7$        & 559.82467   & $26.6 \pm 0.9$        \\
148.97468   & $10.0 \pm 0.7$        & 557.84081   & $15.0 \pm 0.7$        & 559.82577   & $27.1 \pm 0.9$        \\
150.79889   & $33.0 \pm 0.9$        & 557.84309   & $14.0 \pm 0.6$        & 559.82688   & $27.7 \pm 0.9$        \\
150.79997   & $32.3 \pm 0.9$        & 557.84419   & $13.5 \pm 0.6$        & 560.71247   & $40.0 \pm 1.3$        \\
150.80105   & $32.3 \pm 0.9$        & 557.84529   & $13.1 \pm 0.6$        & 560.71357   & $40.5 \pm 1.3$        \\
150.80213   & $31.7 \pm 0.9$        & 557.84639   & $12.6 \pm 0.6$        & 560.71467   & $39.7 \pm 1.3$        \\
153.85512   & $19.0 \pm 0.8$        & 557.84869   & $11.8 \pm 0.6$        & 560.71577   & $39.6 \pm 1.3$        \\
153.85620   & $18.9 \pm 0.8$        & 557.84984   & $11.5 \pm 0.6$        & 560.71862   & $39.1 \pm 1.3$        \\
153.85728   & $19.2 \pm 0.8$        & 557.85100   & $11.1 \pm 0.6$        & 560.71973   & $38.9 \pm 1.3$        \\
153.85835   & $20.0 \pm 0.8$        & 557.85214   & $11.4 \pm 0.6$        & 560.72083   & $39.0 \pm 1.2$        \\
154.70789   & $25.1 \pm 0.9$        & 559.76216   & $7.2 \pm 0.5$         & 560.72193   & $38.7 \pm 1.3$        \\
154.70898   & $25.4 \pm 0.9$        & 559.76327   & $7.3 \pm 0.5$         & 560.72359   & $37.9 \pm 1.2$        \\
154.71008   & $26.1 \pm 0.8$        & 559.76438   & $7.3 \pm 0.5$         & 560.72470   & $37.5 \pm 1.2$        \\
154.71119   & $26.4 \pm 0.8$        & 559.76548   & $7.3 \pm 0.5$         & 560.72580   & $37.2 \pm 1.2$        \\
154.91334   & $23.4 \pm 0.8$        & 559.76721   & $7.5 \pm 0.5$         & 560.72690   & $36.8 \pm 1.2$        \\
154.91444   & $23.6 \pm 0.8$        & 559.76837   & $7.6 \pm 0.5$         & 560.76130   & $21.8 \pm 0.8$        \\
154.91554   & $24.2 \pm 0.8$        & 559.76953   & $7.6 \pm 0.5$         & 560.76242   & $20.8 \pm 0.8$        \\
154.91664   & $24.7 \pm 0.8$        & 559.77068   & $7.9 \pm 0.5$         & 560.76352   & $20.3 \pm 0.8$        \\
156.76397   & $9.5 \pm 0.6$         & 559.77267   & $8.1 \pm 0.5$         & 560.76463   & $19.8 \pm 0.8$        \\
156.76505   & $9.5 \pm 0.6$         & 559.78286   & $10.0 \pm 0.6$        & 563.74274   & $6.3 \pm 0.5$         \\
156.76613   & $10.2 \pm 0.7$        & 559.78396   & $10.3 \pm 0.6$        & 563.74385   & $6.2 \pm 0.5$         \\
156.76721   & $12.1 \pm 0.7$        & 559.78506   & $10.5 \pm 0.6$        & 563.74496   & $7.2 \pm 0.5$         \\
156.91427   & $16.5 \pm 0.8$        & 559.78616   & $10.9 \pm 0.6$        & 563.74607   & $7.3 \pm 0.5$         \\
156.91535   & $16.1 \pm 0.8$        & 559.78814   & $11.5 \pm 0.6$        & 564.84618   & $22.6 \pm 0.8$        \\
156.91643   & $15.6 \pm 0.8$        & 559.78928   & $11.9 \pm 0.6$        & 564.84728   & $23.2 \pm 0.8$        \\
156.91751   & $15.1 \pm 0.7$        & 559.79043   & $12.2 \pm 0.6$        & 564.84839   & $23.7 \pm 0.9$        \\
557.76676   & $39.5 \pm 1.2$        & 559.79158   & $12.7 \pm 0.6$        & 564.84949   & $24.2 \pm 0.9$        \\
557.76787   & $39.7 \pm 1.2$        & 559.79356   & $13.3 \pm 0.6$        & 564.85109   & $24.8 \pm 0.9$        \\
557.76897   & $39.8 \pm 1.3$        & 559.79471   & $13.7 \pm 0.6$        & 564.85220   & $25.4 \pm 0.9$        \\
557.77007   & $39.9 \pm 1.3$        & 559.79586   & $14.1 \pm 0.6$        & 564.85331   & $26.0 \pm 0.9$        \\
            &                       &             &                       & 564.85443   & $26.5 \pm 0.9$        \\ 
\hline=
\end{tabular}
\end{center}
\end{table*}

\section{Space photometric observations}\label{observations}

\subsection{BRITE-Constellation photometry}\label{brite}

{\xic} was observed by BRITE-Constellation \citep{2014PASP..126..573W,2016PASP..128l5001P} during its run in the Canis Major/Puppis~I field between October 26, 2015, and April 18, 2016. The observations were taken by three BRITE satellites, red-filter BRITE-Heweliusz (BHr) and BRITE-Toronto (BTr), and blue-filter BRITE-Lem (BLb), in `chopping mode' \citep{2016PASP..128l5001P}. A short summary of the characteristics of the BRITE data is given in Table \ref{tab:brite}. The photometry was obtained by means of the photometric pipeline described by \cite{2017A&A...605A..26P} and then corrected for instrumental effects according to the procedure described by \citet{2018pas8.conf..175P}. The complete reduced BRITE dataset spans 173.5 days and consists of 152\,112 photometric measurements. 

%Our analysis is based in part on BRITE-Constellation photometry of {\xic} obtained in 2015. Observations were obtained with BRITE-Toronto (BTr; 3 setups, obtaining 64,883 observations over 157 days), BRITE-Lem (BLb; 4 setups, 41,399 observations, 167 days) and BRITE-Heweliusz (BHr; 3 setups, 55,976, 167 days). The observations were reduced and de-trended, yielding typical $1\sigma$ uncertainties of the orbit-averaged measurements of 1.5~mmag for BTr, 3.6~mmag for BHr, and 5.9~mmag for BLb. For this preliminary analysis, we focus on the highest-quality BTr data.

\begin{table}
\centering
\caption{{Space} photometry of {\xic}. RSD and DT stand for residual standard deviation and detection threshold defined as signal-to-noise (S/N) equal to 4 in the frequency spectrum.}
\label{tab:brite}
\begin{tabular}{crrrr}
\hline
Satellite &  \multicolumn{1}{c}{Time}  & \multicolumn{1}{c}{$N_{\rm obs}$} & \multicolumn{1}{c}{RSD} & \multicolumn{1}{c}{DT}\\
ID & \multicolumn{1}{c}{span [d]} & & [mmag] & [mmag] \\
\hline
BLb & 96.7 & 31\,255 & 19.2 & 0.93 \\
BTr & 156.5 & 64\,883 & 5.4 & 0.18 \\
BHr & 166.9 & 55\,974 & 14.4 & 0.53 \\
BTr\,$+$\,BHr & 173.5 & 120\,857 & 10.6 & 0.17 \\
\hline
SMEI & 2884.7 & 25\,581 & 10.6 & 0.32 \\
\hline
{TESS} & {21.8} & {14\,814} & {0.2} & {0.03} \\
\hline\hline
\end{tabular}
\end{table}

\subsection{SMEI photometry}\label{smei}
The Solar Mass Ejection Imager (SMEI) experiment \citep{2003SoPh..217..319E,2004SoPh..225..177J} was placed on-board of the Coriolis spacecraft and was aimed at measuring sunlight scattered by free electrons of the solar wind. We used photometry of {\xic} obtained between 2003 and 2010 and available through the University of California San Diego (UCSD) web page\footnote{http://smei.ucsd.edu/new\_smei/index.html}. The SMEI time series are affected by long-term calibration effects, especially a repeatable variability with a period of one year. {The raw SMEI UCSD photometry of {\xic} were corrected for the one-year variability by subtracting an interpolated mean light curve, which was obtained by folding the raw data with the period of one year, calculating median values in 200 intervals in phase, and then interpolating between them. In addition, the worst parts of the light curve and outliers were removed. The data points were also assigned individual uncertainties calculated using the scatter of the neighbouring data. Then, a model consisting of the dominant pulsation frequency (4.77~d$^{-1}$) and its detectable harmonics was fitted to the data. Finally, the low-frequency instrumental variability was filtered out by subtracting a trend using residuals from the fit. The last two steps were iterated several times.} 
%\comment{APi, could you please reformulate the previous sentence? Its meaning wasn't clear to some readers.} 
The SMEI dataset that we analysed spans 2885 days and consists of 25\,581 photometric measurements.

\subsection{TESS photometry}\label{tess}
{The primary goal of the NASA's TESS mission \citep{2014SPIE.9143E..20R,2015JATIS...1a4003R} is the detection of planets by means of the transit method. TESS observations cover almost the entire sky, excluding only the regions with low Galactic latitudes ($|b|<$~6$\degr$). Observations are carried out with a 30-min cadence, but selected stars, including {\xic}, are observed with a shorter, 2-min cadence. The star was observed with TESS camera \#2 in Sector 6. The observations spanned 21.8~d between December 11, 2018, and January 7, 2019, and consisted of 15\,678 data points. In the subsequent analysis (Sect.~\ref{freqanalysis}) we used SAP fluxes and removed all data points with quality flag different from 0.}
%\subsection{Ground-based Str\"omgren photometry}
%
%\modif{Add/discuss Gerald's photometry here?}

\subsection{Frequency analysis}\label{freqanalysis}
\begin{figure*}
  \centering
    \includegraphics[width=9cm]{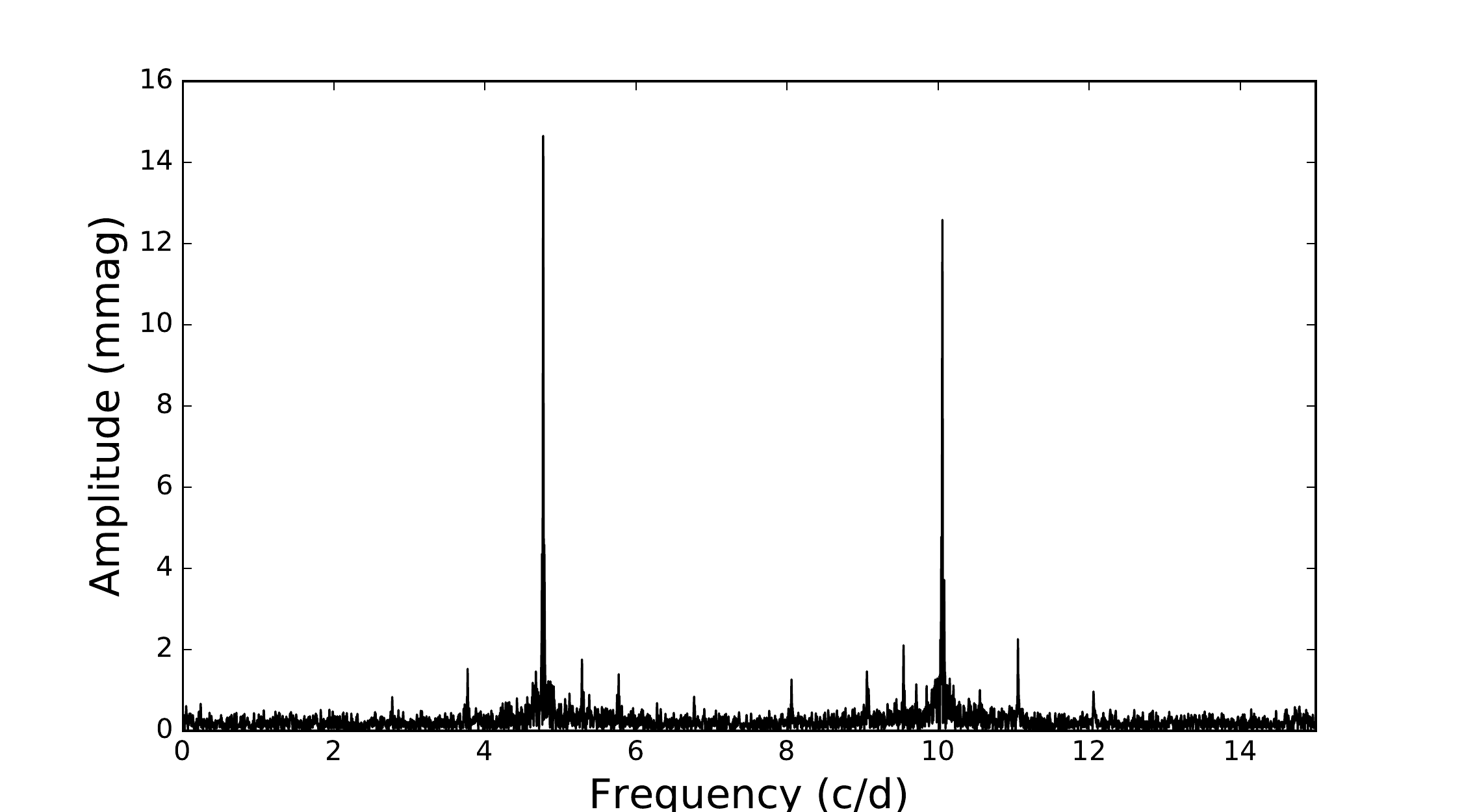} \hspace{-0.5cm}\includegraphics[width=9cm]{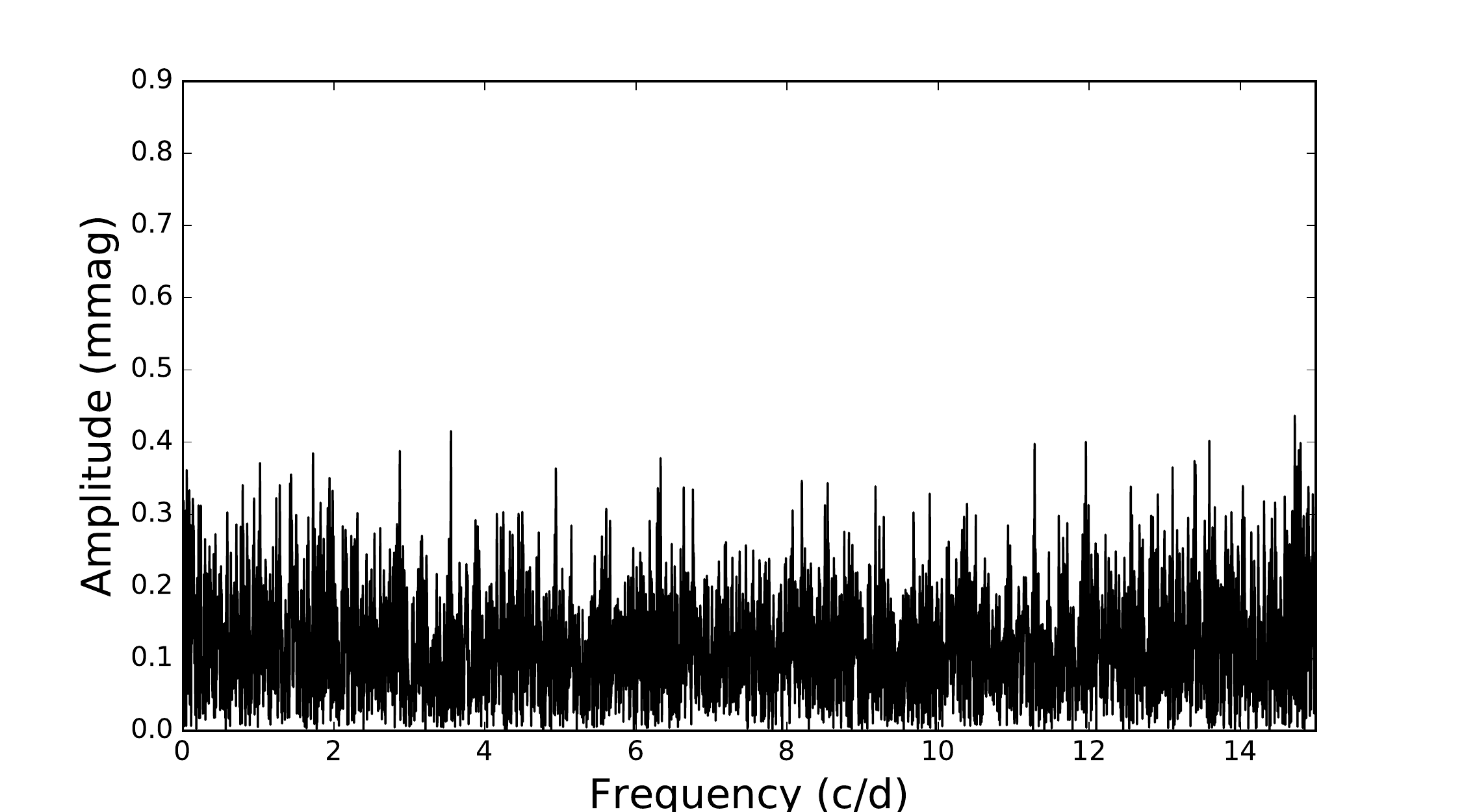}
        \includegraphics[width=9cm]{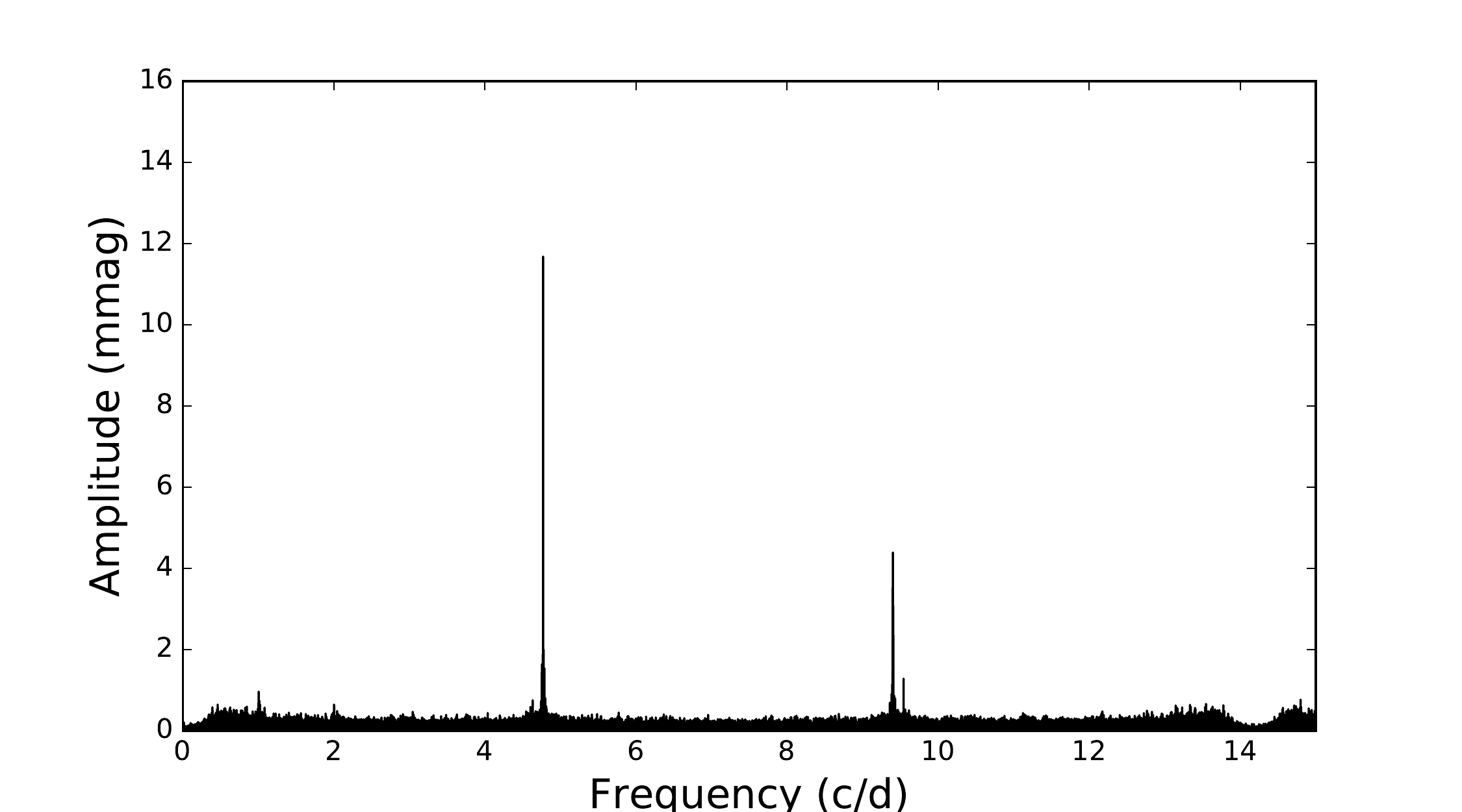}\hspace{-0.5cm}\includegraphics[width=9cm]{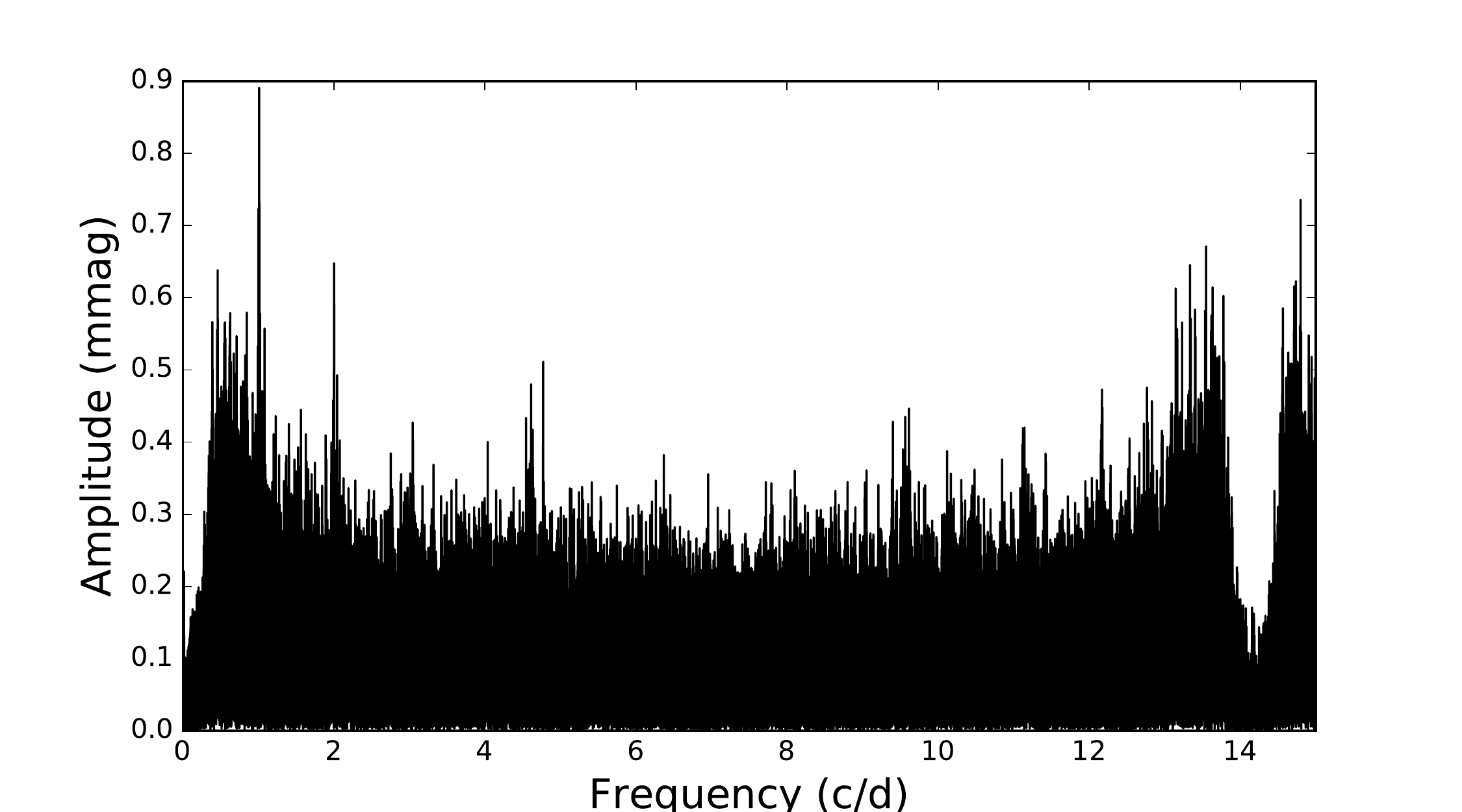}
        \includegraphics[width=9cm]{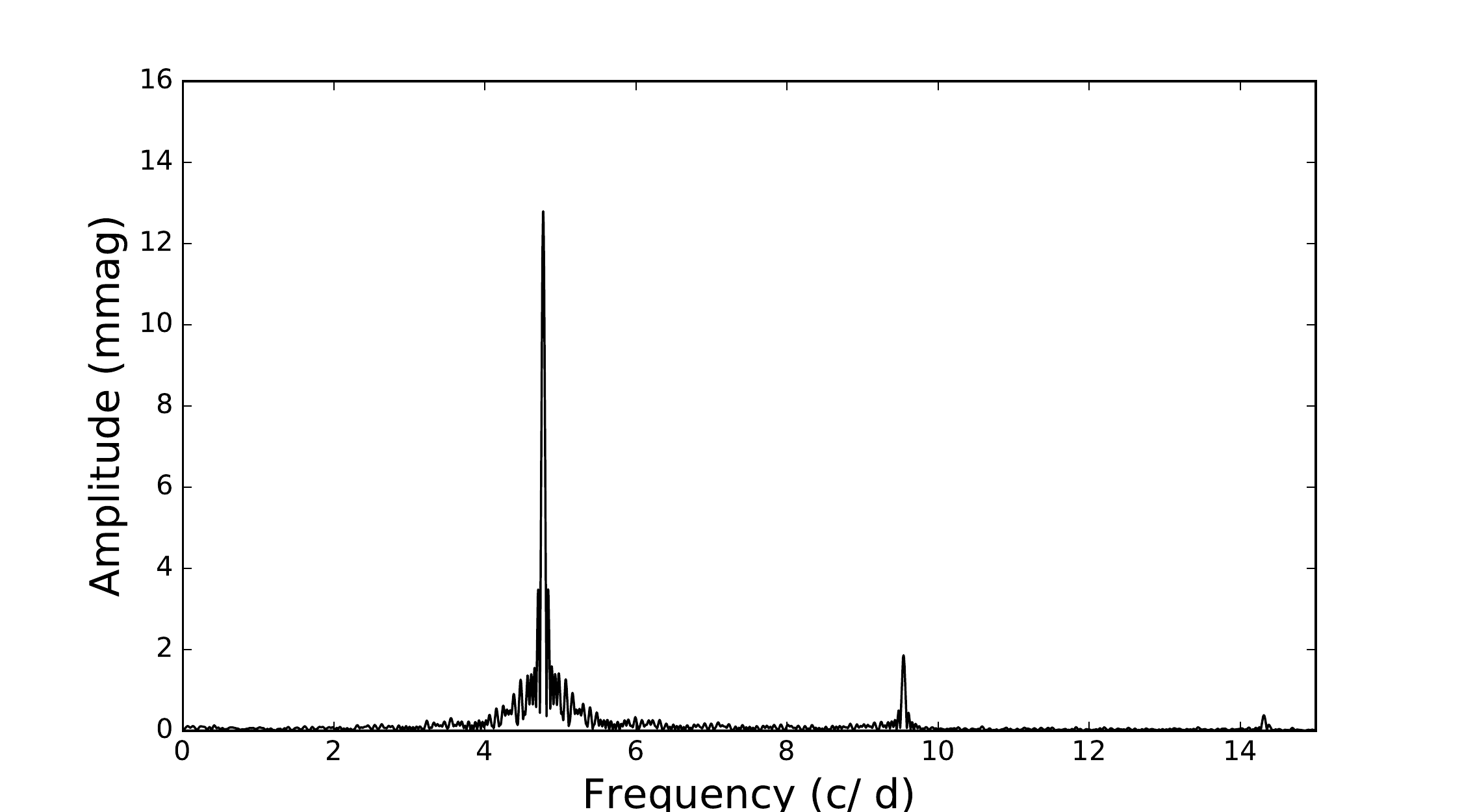}\hspace{-0.5cm}\includegraphics[width=9cm]{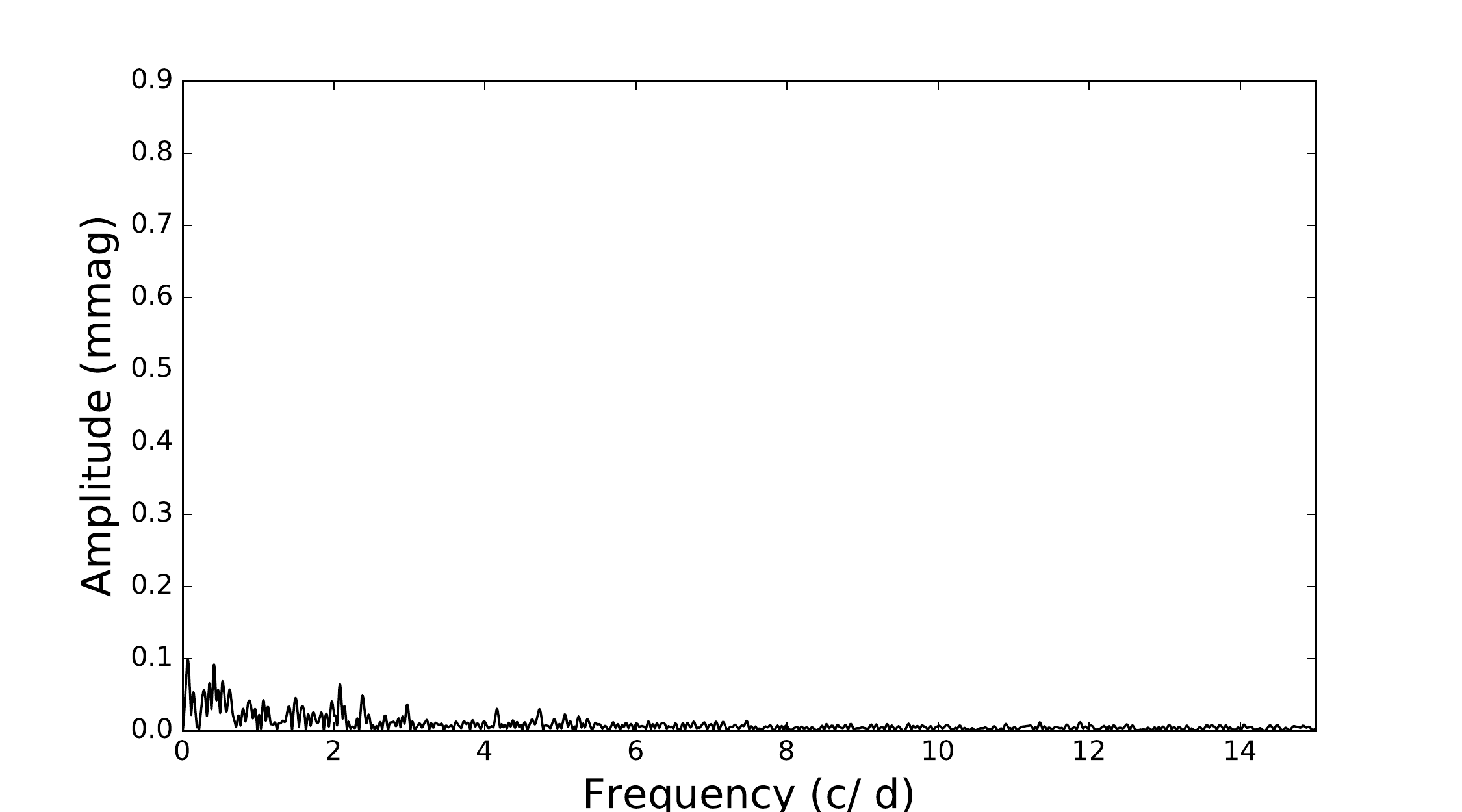}
\caption{Fourier amplitude spectra of BTr+BHr data (upper frames) and SMEI data (lower frames) in mmag. {\em Upper left:} BTr+BHr data. {\em Upper right:} Spectrum of residuals following prewhitening with the fundamental frequency of 4.771491~d$^{-1}$ and the first two harmonics. {\em Middle left:} Orbit-averaged SMEI data. {\em Middle right:} Spectrum of residuals following prewhitening with the fundamental frequency of 4.771517~d$^{-1}$ and its first two harmonics. {\em Bottom left:} TESS data. {\em Lower right:} Spectrum of residuals following prewhitening with the fundamental frequency of 4.771483(6)~d$^{-1}$ and its first four harmonics. In the BRITE and SMEI residuals, the peaks at 1 and 2~d$^{-1}$ are instrumental, as are peaks between 3 and 5~d$^{-1}$. }
    \label{fig1}
\end{figure*}

Fourier analysis with prewhitening was performed on the BRITE photometry using the {\sc Period04} package \citep{2005CoAst.146...53L}.  We combined the BTr and BHr data and analyzed them as a single dataset. One significant frequency was detected at 4.771491(7)~d$^{\rm -1}$ (amplitude of 14.8~mmag, corresponding to a period of $0.2095781(3)$~d), along with its first two harmonics (with amplitudes of 2.0 and 0.6~mmag, respectively). The original and prewhitened (up to the second harmonic) Fourier amplitude spectra of the red BRITE data are illustrated in Fig.~\ref{fig1}. As is evident in Fig.~\ref{fig1}, the BRITE data reveal no evidence for new independent pulsation frequencies with amplitudes larger than about 0.17~mmag. Analysis of the blue BRITE data yields compatible results, but with higher uncertainties and upper limits (0.93~mmag).

%BRITE amplitudes
%Fundamental = 14.818805
%1st Harmonic = 1.972953
%2nd Harmonic = 0.631106

%4.7713981, 9.543084, 14.3147707 

A similar analysis was performed on the SMEI data (Fig.~\ref{fig1}). The fundamental pulsation frequency 4.771517(2)~d$^{\rm -1}$ (amplitude of 11.7~mmag, corresponding to a period of $0.2095770(1)$~d) was clearly detected. No significant non-instrumental signal was detected after prewhitening with the fundamental frequency and its first two harmonics (amplitudes of 1.5~mmag and 0.9~mmag, respectively). However, the fundamental pulsation frequency determined using the SMEI data is somewhat higher than that derived using the BRITE data. This difference is consistent with the reported evolution of the pulsation period of \xic, and will be discussed further below. The detection threshold of the SMEI data, taking into account uncertainties associated with the removal of long-term trends, is about {0.3 mmag}.  %\comment{BP asked, "How significant is this difference? SMEI doesn't have a filter so how does its corresponding transmission function compare with the BRITE red filter?" GAW replies: "The significance is reflected in the uncertainties. It's not clear to me why the measured period would be a function of the passband details.} \comment{APi: SMEI passband is detemined primarily by the sensitivity of CCD. I see no reason why period should depend on passband. The periods are different because the data were obtained at different times. In the presence of period changes this results in the difference.}

\begin{figure}
  \centering
    \includegraphics[width=9cm]{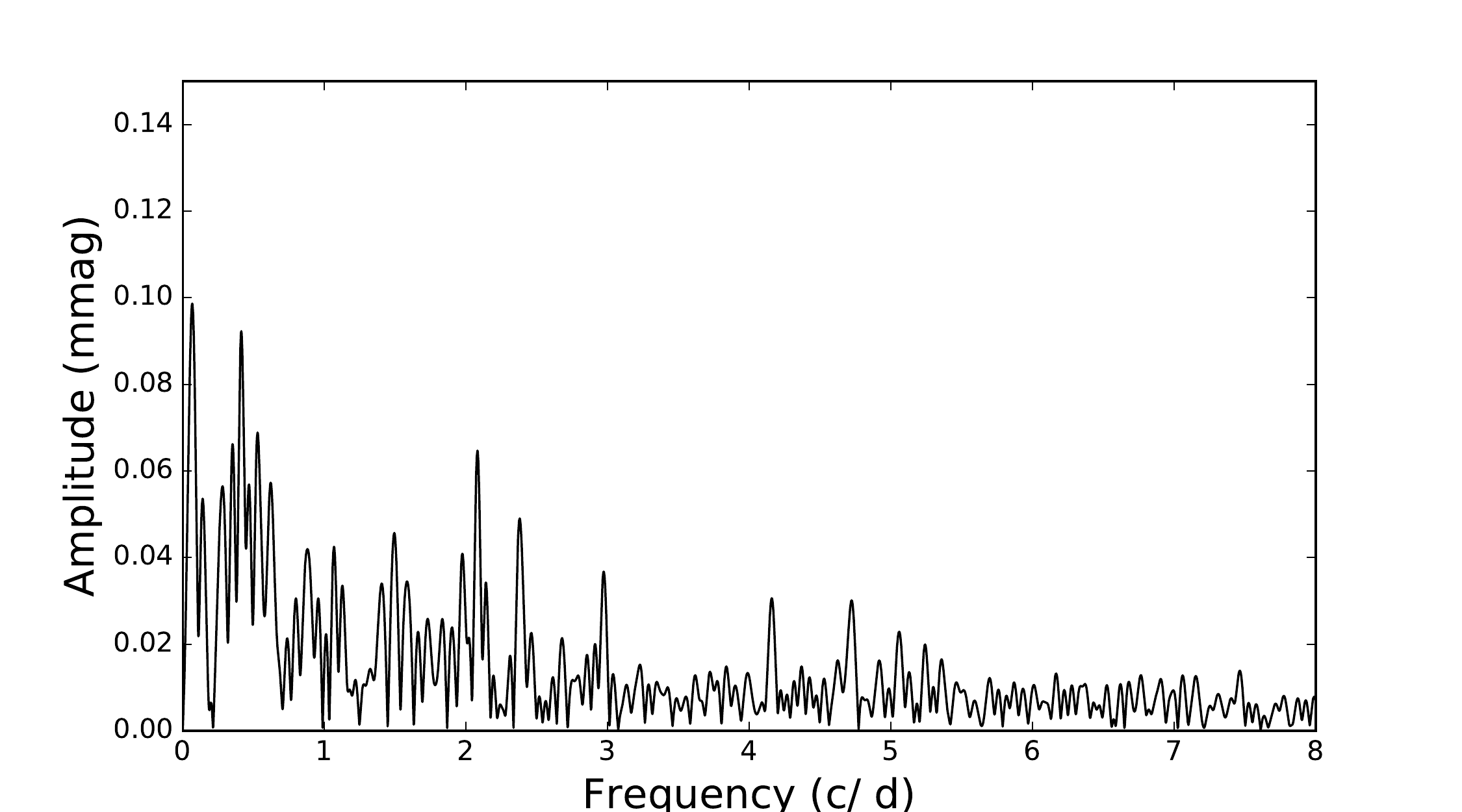}
\caption{Fourier amplitude spectra of TESS sector 6 photometry showing weak peaks at low frequencies.}
    \label{fig1b}
\end{figure}

{Finally, we analyzed TESS photometry of {\xic}. With the exceptionally low detection threshold of about 0.03~mmag (Table \ref{tab:brite}) we were able to detect not only the dominant frequency (at {4.771483(6)}~d$^{\rm -1}$, with an amplitude of {12.8}~mmag), but also its four lowest harmonics.  In addition (Fig.\,\ref{fig1b}), the frequency spectrum shows extra power below $\sim$5~d$^{-1}$. Several significant peaks with amplitudes below 0.12~mmag can be identified. They may correspond to both $p$ and/or $g$ modes. The detailed analysis and possible seismic modeling with the use of these frequencies is, however, beyond the scope of this paper.}

%\comment{Andrzej, should this be three or four? APi: Four harmonics stand clearly above the noise, but the highest, 5$f_1 =$ 23.857~d$^{-1}$, has an amplitude of 0.017~mag, that is, below 0.03~mmag. The problem is that the detection threshold is calculated from residuals in which the low-frequency signal is still present and in the range 0\,--\,30~d$^{-1}$. That's why we get 0.03~mag, though the detection threshold calculated from the same data in the range 10\,--\,40~d$^{-1}$, which would be more appropriate for 5$f_1$, is equal to 0.0125~mmag. With this detection threshold, the S/N for $5f_1$ is therefore equal to 5.51.}

%SMEI amplitudes
%Fundamental = 11.708245
%1st Harmonic = 1.475242
%2nd Harmonic = 0.890871

%Here's the updated figures!
%
%The first is the original plot with zero pre whitening done (TESS_orig). The second is having the third harmonic pre whitened (TESS_3), the third is the same as the second but cut from 0-8 c/d (TESS_3_cut). I changed the y-axis compared to the BRITE SMEI pre whitened files (they were at 0.9) since the signal is down at 0.1, but if you'd prefer it to match, that's an easy fix.
%
%If you need the amplitudes and frequencies:
%
%Fundamental: Amp = 12.8, Freq. = 4.772105
%1st: Amp = 1.81, Freq. = 9.543875
%2nd: Amp = 0.36, Freq. = 14.313684
%3rd: Amp = 0.11, Freq. = 0.215870
%
%I can do a further prewhiten if you/Andzrej deems necessary, let me know!

%\comment{Tony asks: [In the upper left panel I see a peak at the fundamental frequency and weakly at the 1st harmonic (9.54 c/d) but nothing at the 2nd harmonic (14.31 c/d), contrary to this statement. Contrary to which statement? The fact that you can't see them in this picture doesn't imply that they aren't there - they're just very weak, but still significant.} \comment{Tony asks: And what are the other 8-9 peaks at various frequencies here, some higher than the 2nd harmonic? Clearly elements of the window function, as they disappear with prewhitening.}

%\begin{figure}
%  \centering
%    \includegraphics[width=9cm]{BRITE_binned.pdf}
%    \caption{The BRITE red (combined BTr and BHr) and blue (BLb) data phased according to the ephemeris \comment{[Sean, which ephemeris did you use to phase the data? The y-axis should read "Magnitude (mmag)"]}. The data have been binned to a resolution of 0.01 cycles. \modif{Recompute binned data using weights. Add SMEI. Increase x-range to -0.25 to 1.25.}}
%    \label{fig_lightcurve}
%\end{figure}

%\begin{figure}
%  \centering
%    \includegraphics[width=8cm]{Freq/Smei_prewhite_fun_15.pdf}
%    \includegraphics[width=8cm]{Freq/Smei_prewhite1_15_II.pdf}
%    \caption{Fourier amplitude spectra of SMEI data (in mmag). {\em Top -}\ Orbit-averaged data. {\em Upper right -}\ Spectrum of residuals following prewhitening with the fundamental frequency of 4.771515~d$^{\rm -1}$. {\em Middle -}\ Spectrum of residuals following prewhitening with the first harmonic frequency. {\em Bottom -}\ Spectrum of residuals following prewhitening with the second harmonic frequency of XXX. \comment{Is this an optimal representation of the data?}}
%    \label{fig2}
%\end{figure}

\section{Spectropolarimetric observations and radial velocities}\label{specpol}

In addition to the ESPaDOnS RVs published by \cite{2017MNRAS.471.2286S}, we have included new RV measurements obtained from follow-up ESPaDOnS observations in 2018 and 2019\footnote{Program codes and 18AC19 and 19AC20.}. Eight spectropolarimetric sequences were obtained in 2018; the magnetic analysis of these data were described by \cite{2018MNRAS.478L..39S}. A further 19 sequences were obtained in 2019; the magnetic analysis will be presented by Erba et al.\ (in prep.).

ESPaDOnS is an echelle spectropolarimeter with a high resolving power ($\lambda/\Delta\lambda \sim 65,000$), with a wavelength range of about 370 nm to 1000 nm, mounted at the Cassegrain focus of the 3.6 m Canada-France-Hawaii Telescope (CFHT). The instrument properties and data reduction were described in detail by \cite{2016MNRAS.456....2W}. Each spectropolarimetric sequence consists of 4 differently spectra. The 2019 observations are essentially identical to the 2018 observations described by \cite{2018MNRAS.478L..39S}, with a mean peak signal-to-noise (S/N) per spectral pixel of about 400 in the individual intensity spectra. In the case of $\xi^1$ CMa the exposure time per individual spectrum (72 s) is a much lower fraction of the pulsation period than the combined 8 minute exposure-plus-readout time of a full sequence (0.004\% vs.\ 0.027\%), therefore individual spectra were used for RV measurements, thus yielding 32 measurements in 2018 and 76 measurements in 2019. RVs were measured from the weighted means of the centres-of-gravity across multiple unblended spectral lines, using the same method and line list described by \cite{2017MNRAS.471.2286S}. 

\section{Evolution of the pulsation period from 1906-2017}

To investigate the behaviour of the pulsation period of \xic\ we have constructed an {\oc} diagram using all available spectroscopic and photometric observations. Since both light and radial velocity can be described by a single periodicity, the times of maximum light (and radial velocity) were derived by fitting a function of the form
\begin{equation}
\sum\limits_{m=1}^N A_m\sin(2\pi mft + \phi_m),
\label{eq:tsin}
\end{equation}
to the light or radial velocity time-series. In this equation, $f$ stands for the pulsation frequency, $t$ is the time elapsed from the initial epoch, while $A_m$ and $\phi_m$ are respectively semi-amplitudes and phases of the consecutive harmonic terms. Depending on the data, the fitted model included all detectable harmonics, up to $N=5$. The harmonics account for deviations from the sinusoidal shape of the light or radial velocity curve. The times of maximum summarized in {Tables \ref{tab:tmax-rvel} and \ref{tab:tmax-phot}} correspond to the maximum of the fit given by Eq.~(\ref{eq:tsin}), that is, including all detectable harmonics. 
In our analysis, all dates are given as HJD at the mid-time of exposures. 

 \begin{table*}
\centering
\caption{Times of maximum radial velocity for {\xic}. {Columns give HJD of maximum {RV}, the number of cycles before/since the reference ephemeris, the inferred O-C, the source of the data, the number of observations, and any notes or comments.}}
\label{tab:tmax-rvel}
\begin{tabular}{lrccrl}
\hline
\multicolumn{1}{c}{$T_{\rm max}-$}& $E$ & (O$-$C) & \multicolumn{1}{c}{Source of} & $N_{\rm obs}$ & Notes, comments\\
\multicolumn{1}{c}{HJD\,2\,400\,000}&&[d] &\multicolumn{1}{c}{data}& \\
\hline
17221.336(10)& $-$114874 & $+$0.06029 & \cite{1926ApJ....64....1F} & 5 & 1905 -- 1906\\
19523.709(10)& $-$103888 & $+$0.03734 & \cite{1928PLicO..16....1C} & 5 & 1909 -- 1913\\
22697.054(10)& $-$88746 & $-$0.01448 & \cite{1921PDO.....5...45H} & 7 & best of 1921 only\\
34439.0916(11)& $-$32718 & $-$0.06859 & \cite{1955ApJ...122...95M} & 45 & 1952 -- 1953\\
34816.1163(22)& $-$30919 & $-$0.07022 & \cite{1956PASP...68..263M} & 18 & 1954\\
51624.74742(39)& 49284 & $-$0.02292 & \cite{2017MNRAS.471.2286S} & 51 &  Feb/Apr 2000\\
51888.60434(22)& 50543  & $-$0.02156 & \cite{2017MNRAS.471.2286S}  & 52 & Dec 2000\\
51949.17159(23)& 50832 & $-$0.02163 & \cite{2017MNRAS.471.2286S} & 51 & Feb 2001\\
52233.98574(24)& 52191 & $-$0.02058 & \cite{2017MNRAS.471.2286S} & 58 & Nov 2001\\
52580.41476(73)& 53844 & $-$0.01986 & \cite{2017MNRAS.471.2286S} & 10 & Oct/Nov 2002\\
52989.29838(24)& 55795 & $-$0.01804 & \cite{2017MNRAS.471.2286S} & 71 & Dec 2003\\
53078.99670(25)& 56223 & $-$0.01804 & \cite{2017MNRAS.471.2286S} & 60 & Mar 2004\\
53278.30372(30)& 57174 & $-$0.01732 & \cite{2017MNRAS.471.2286S} & 39 & Sep/Oct 2004\\
55180.42624(24)& 66250 & $-$0.00204 & \cite{2017MNRAS.471.2286S} & 79 & Jan 2008 -- Dec 2010\\
56308.58266(25)& 71633 & $-$0.00947 & \cite{2017MNRAS.471.2286S} & 56 & Feb 2012 -- Jan 2014\\
57805.59816(24)& 78776 & $+$0.02717 & \cite{2017MNRAS.471.2286S} & 85 & Feb/Mar 2017\\
{58560.28537(11)}& {82377} & {$+$0.03301} & {This paper} & {77} & {Mar 2019}\\
\hline\hline
\end{tabular}
\end{table*}

\subsection{Radial velocity data}
The radial velocity data consist of a rich, high-quality data set of spectroscopic measurements obtained in the years {2000\,--\,2019} and five archival data sets, which extend the study of period changes to over a century. The  2000\,--\,2016 spectroscopy was used by \citet{2017MNRAS.471.2286S} and \citet{2018pas8.conf..154B} to conclude that the pulsation period of {\xic} changes with the constant rate of $+$0.9 $\pm$ 0.1\,s/cen. The archival data include radial velocities published by \cite{1926ApJ....64....1F} (these are the corrected discovery data of \cite{1907ApJ....25R..59F} plus one additional spectrum), \cite{1921PDO.....5...45H}, \cite{1928PLicO..16....1C}, and \cite{1955ApJ...122...95M,1956PASP...68..263M}. All data are available and were used to derive the times of maximum presented in Table \ref{tab:tmax-rvel}. Heliocentric corrections were applied to all data for which the reported time was given in Julian Days. The 2000\,--\,2016 spectroscopy was split into 11 subsets, usually corresponding to a single observing season. In case the number of observations was small, data from adjacent seasons were combined.% \comment{Check if information about the new RV measurements needs to be added here.}

\subsection{Photometric data}
The archival photometry of {\xic} includes ground-based observations of \cite{1954PASP...66..200W}, \cite{1962ZA.....56..141V}, \cite{1971ApJ...170..345W}, \cite{1973MNRAS.162...25S}, and \cite{1992AAS...96..207H}. In addition, we used Str\"omgren $uy$ photometry obtained at Fairborn Observatory in 2018 by one of us (G.H.). Surprisingly, the star was frequently observed from  space. The data sets we used include ultraviolet (UV) photometry from the TD-1A \citep{1977AA....61..815B,1980AAS...39..301B} and ANS \citep{1979AA....79..115L} satellites and the optical-domain data from Hipparcos, BRITE (Sect.~\ref{brite}), SMEI (Sect.~\ref{smei}), {and TESS (Sect.~\ref{tess}).} %\comment{BP notes that there are also Kamogata Wide Field Survey two-band data available from JD 2455551.14287 to 2458561.923634. APi: I was not aware that such data exist. I have checked them. The $V$ and $I_{\rm C}$ data cover 9 and 6 seasons, respectively. Only $f_1$ is detected in these data sets with amplitudes 18.1 and 14.2~mmag, residual standard deviations of 41.8 and 34.4~mmag, and detection thresholds, 12.6 and 11.1~mmag, respectively. This means that if we would like to split the data for seasons, we will not detect even $f_1$. Since we have precise data for the epochs covering these observations, I propose not to use them --- they are very poor and will not change anything.}

\subsubsection{Effective wavelengths}
In the presence of phase lags between the times of maximum in different photometric bands (Sect.\,\ref{plags}), it became necessary to derive effective wavelengths for the passbands used in the observations of {\xic}. They were defined with the following expression:
\begin{equation}
\lambda_{\rm eff} = \frac{\int\limits_{\lambda_1}^{\lambda_2} \lambda S(\lambda) T_1(\lambda) T_2(\lambda)T_3(\lambda)d\lambda}{\int\limits_{\lambda_1}^{\lambda_2} S(\lambda) T_1(\lambda) T_2(\lambda)T_3(\lambda)d\lambda},
\end{equation}
where $S(\lambda)$ represents a model spectrum with $T_{\rm eff}= 27500$\,K, $\log g=3.75$ taken from the OSTAR2002 grid of models \citep{2003ApJS..146..417L}. The model parameters are close to the values derived for {\xic} by \cite{2017MNRAS.471.2286S}.  
%The effective wavelengths can remaby anchoring a black-body function with $T=T_{\rm eff}$ at $\lambda=$320~nm.  
The variables $T_1(\lambda)$, $T_2(\lambda)$, and $T_3(\lambda)$ (all included optionally) are filter transmission curves, detector sensitivity curves, and (for ground-based observations) the atmosphere transmission curves. Values of $\lambda_1$ and $\lambda_2$ were chosen to encompass the non-zero values of the sensitivity and transmission curves. Details concerning $T_1$, $T_2$, and $T_3$ are as follows: 

%The model was extended to UV ($\lambda <$ 320~nm) \comment{[GAW comments: ``The BSTAR spectra are computed at wavelengths shorter than 320 nm. Do you intend to make any change to this analysis?" Andrzej replied ``Right, I overlooked this. I can correct this although the UV effective wavelengths are not used, so in principle we can remove them from Table 3." GAW asks ``Sorry, I don't understand. Why aren't they used?"]} \modif{The values in in the table, but since they aren't used no update is needed.}
%\comment{[Matt comments: "Probably won't effect the results, but I found 27 kK to be a better fit."] Andrzej replies "I have checked this. As can be expected, the values of effective wavelengths are shifted redwards by 2 - 9 nm, depending on the passband. Do you want me to recalculate this?"]} 

\begin{itemize}
\item \cite{1973MNRAS.162...25S}: $T_1$ was Johnson $V$ (the author used Corning 3384 filter which defines $V$ band and an EMI\,8094S photomultiplier with a S-11 photocathode, similar to that used in a 1P21 photmultiplier, which defines $V$), that is, a combination of filter transmission and detector sensitivity, taken from ADPS\footnote{http://ulisse.pd.astro.it/Astro/ADPS/} \citep{2000A&AS..147..361M}. For $T_2$ we took the extinction curve (twice as large as in La Silla; see Geneva photometry below). The same combination of $T_1$ and $T_2$ was adopted for the data published by \cite{1954PASP...66..200W} and  \cite{1962ZA.....56..141V}.
\item \cite{1971ApJ...170..345W}: $T_1$ was Newell $v$ band transmission taken from ADPS, $T_2$ was the same as for  \cite{1973MNRAS.162...25S}. 
\item TD-1A: $T_1$: Passband centres and effective widths were taken from the HEASARC web page\footnote{https://heasarc.gsfc.nasa.gov/W3Browse/all/td1.html}; shapes were approximated by an $\exp(-(\Delta\lambda/\sigma)^4)$ function. $T_2$, corresponding mainly to the detector response, was estimated for each TD-1A passband from fig.~8 of \cite{1973MNRAS.163..291B}.
\item ANS: $T_1$: Instrument response was taken from \cite{1975A&A....39..159V}.
\item Geneva: Filter transmission curves $T_1$ were taken from ADPS; $T_2$ was S-11 photocatode QE curve\footnote{http://www.r-type.org/pdfs/9531.pdf}. As $T_3$, the La Silla extinction coefficient dependence was taken\footnote{https://www.eso.org/sci/observing/tools/Extinction.html}.
\item Hipparcos: $T_1$: The passband as defined by \cite{2000PASP..112..961B} was used. 
\item SMEI: For $T_1$ we adopted the typical E2V Technologies standard front-illuminated CCD sensitivity curve\footnote{https://www.e2v.com/content/uploads/2017/08/ccdtn101.pdf} because it seems to be similar to the description of the E2V CCD05-30-231A chip, given by \cite{2003SoPh..217..319E}.
\item BRITE: $T_1$: The BRITE filter transmission curves from \cite{2014PASP..126..573W}. $T_2$:  Kodak KAI-11002 sensitivity curve from the product sheet\footnote{http://www.onsemi.com/pub/Collateral/KAI-11002-D.PDF}.
\item Fairborn Observatory ground-based observations in Str\"omgren $u$ and $y$ filters. For $T_1$ the transmissions curves for Str{\"o}mgren $u$ and $y$ filters from ADPS were taken. For $T_2$ and $T_3$ the QE curve from \cite{1997PASP..109..697S} for the Thorn-EMI 9124QB photomultiplier and the La Silla extinction were used, respectively.
\item {TESS: The TESS curve, including both the sensitivity of the detector \citep{2015JATIS...1a4003R} and the filter transmission curve, has been taken from the NASA's High Energy Astrophysics Science Archive Research Center web page\footnote{https://heasarc.gsfc.nasa.gov/docs/tess/data/tess-response-function-v1.0.csv}.}
\end{itemize}
The calculated effective wavelengths are given in Table \ref{tab:tmax-phot}.
\begin{table*}
\centering
\caption{\label{tab:tmax-phot}Times of maximum light and related information for {\xic}. Columns give HJD of maximum light, correction to the HJD of maximum light in the visual domain {(Eq.~\ref{eq-corr})}, the number of cycles before/since the reference ephemeris, the inferred O-C, the passband, the effective wavelength, the source of the data, the number of observations, and any notes or comments.}
\setlength\tabcolsep{3pt}
\begin{tabular}{lcrccclrl}
\hline
\multicolumn{1}{c}{$T_{\rm max}^{\rm obs}-$}& $C_{\rm Vis}$ &$E$ & (O$-$C) & Passband(s) & $\lambda_{\rm eff}$  & \multicolumn{1}{c}{Source of} & $N_{\rm obs}$ & Notes, comments\\
\multicolumn{1}{c}{HJD\,2\,400\,000}&[d]&&[d] &&[nm]&\multicolumn{1}{c}{data}& \\
\hline
34719.3697(38)&$-$0.0019& $-$31381 & +0.0052 & yellow & 547 & \cite{1954PASP...66..200W} & 99 & data read off the figures\\
37658.0342(15) &$-$0.0019& $-$17359 & $+$0.0020 & $Y$ &  547 & \cite{1962ZA.....56..141V} & unkn. & combined 13 $T_{\rm max}$\\
40562.9555(13) &$-$0.0013& $-$3498 & $-$0.0021 & Newell $v$& 533& \cite{1971ApJ...170..345W} &unkn.& from  \cite{1973MNRAS.162...25S}\\
41296.0514(6)&$-$0.0019& 0 & $-$0.0019 & yellow & 547 & \cite{1973MNRAS.162...25S} & 562 & original $T_{\rm max}$\\
%41660.075(7) &---& 1737 & $-$0.0095 & 155 & 151 & \cite{1980AAS...39..301B} & 8 & TD-1A\\
%41660.078(8) &---& 1737 & $-$0.0064 & 195 & 196 & \cite{1980AAS...39..301B} & 8 & TD-1A\\
%41660.082(12) &---& 1737 & $-$0.0021 & 235 & 235 & \cite{1980AAS...39..301B} & 8 & TD-1A\\
%41660.080(32) &---& 1737 & $-$0.0040 & 275 & 273 & \cite{1980AAS...39..301B} & 8 & TD-1A\\
{41406.911(2)} &{---}& {529} & {$-$0.0058} & {155\,--\,275} & {214} & \cite{1977AA....61..815B} & {8} & {TD-1A}\\
42867.8747(40) &---& 7500 & $+$0.0071 & 155, 180, 220 & 184 & \cite{1979AA....79..115L} & 25 & ANS\\
42867.8727(40) &---& 7500 & $+$0.0051 & 250, 330 & 289 & \cite{1979AA....79..115L} & 25 & ANS\\
47504.3265(10)&$+$0.0061& 29623 & $+$0.0262 & Geneva $U$ & 348 & \cite{1992AAS...96..207H} & 203 & \\
47504.3292(11)&$+$0.0040& 29623 & $+$0.0268 & Geneva $B_1$ & 401 & \cite{1992AAS...96..207H} & 203 & \\
47504.3300(10)&$+$0.0032& 29623 & $+$0.0268 & Geneva $B$ & 419 & \cite{1992AAS...96..207H} & 203 & \\
47504.3305(12)&$+$0.0022& 29623 & $+$0.0263 & Geneva $B_2$ & 446 & \cite{1992AAS...96..207H} & 202 & \\
47504.3342(17)&$-$0.0013& 29623 & $+$0.0265 & Geneva $V_1$ & 534 & \cite{1992AAS...96..207H} & 204 & \\
47504.3326(19)&$-$0.0014& 29623 & $+$0.0248 & Geneva $V$ & 535 & \cite{1992AAS...96..207H} & 202 & \\
47504.3359(18)&$-$0.0028& 29623 & $+$0.0267 & Geneva $G$ & 570 & \cite{1992AAS...96..207H} & 200 & \\
48380.3596(13)&$+$0.0004& 33803 & $+$0.0280 & $H_{\rm p}$ & 490 & Hipparcos & 216 & \\
52992.3066(7)&$-$0.0046& 55809 & $+$0.0515 & SMEI & 615 & SMEI (UCSD) & 5116 & SMEI data, part 1\\
53592.1155(6)&$-$0.0046& 58671 & $+$0.0553 & SMEI & 615 & SMEI (UCSD) & 5116 & SMEI data, part 2\\
54123.6040(8)&$-$0.0046& 61207 & $+$0.0604 & SMEI & 615 & SMEI (UCSD) & 5116 & SMEI data, part 3\\
54656.5586(7)&$-$0.0046& 63750 & $+$0.0645 & SMEI & 615 & SMEI (UCSD) & 5116 & SMEI data, part 4\\
55252.1769(7)&$-$0.0046& 66592 & $+$0.0692 & SMEI & 615 & SMEI (UCSD) & 5117 & SMEI data, part 5\\
57367.6476(3)&$+$0.0030& 76686 & $+$0.0924 & BRITE blue & 424 & this paper & 31255 & BLb \\
57384.8364(11)&$-$0.0019& 76768 & $+$0.0911 &Str{\"o}mgren $y$& 546 &this paper&494& APT (Fairborn)\\
57385.0417(5)&$+$0.0060& 76769 & $+$0.0947  &Str{\"o}mgren $u$& 345 &this paper&501& APT (Fairborn)\\
57401.60040(10)& $-$0.00415& 76848 & $+$0.08683 & BRITE red & 604 & this paper & 120857 & BHr + BTr \\
{58479.250031(7)}& {---} & {81990} & {$+$0.103386} & {TESS} & {733} & {TESS} & {14812} & {Sector 6, camera \#2}\\
\hline\hline
\end{tabular}
\end{table*}
\begin{figure}
\centering
\includegraphics[width=\columnwidth]{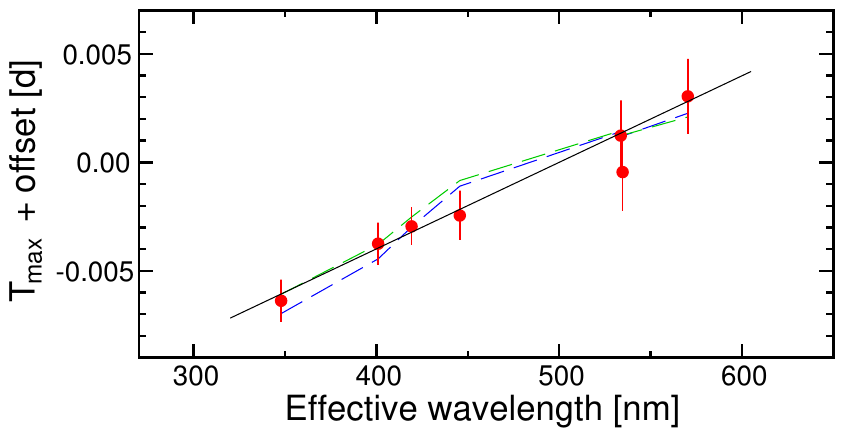}
\caption{The times of maximum light derived from the Geneva photometry \citep{1992AAS...96..207H} in the visual domain (red points). The linear coefficient of the fitted line corresponds to $A_{\rm Vis}= (+$3.99\,$\pm$\,0.33)\,$\times$\,10$^{-5}$~d\,(nm)$^{-1}$. {The dashed lines represent the theoretical dependences for the fundamental radial mode in two BG models with $M=14$\,$M_\odot$, $T_{\rm eff}=27$\,kK, $\log g = 3.74$, and $\xi=2$\,{\kms} (blue) and 10\,{\kms} (green); see Sect.\,\ref{seisminf} for explanation.}}
\label{fig:pl}
\end{figure}

\subsubsection{Correction for the phase lag between photometric bands}\label{plags}
\begin{figure*}
\centering
\includegraphics[width=0.9\textwidth]{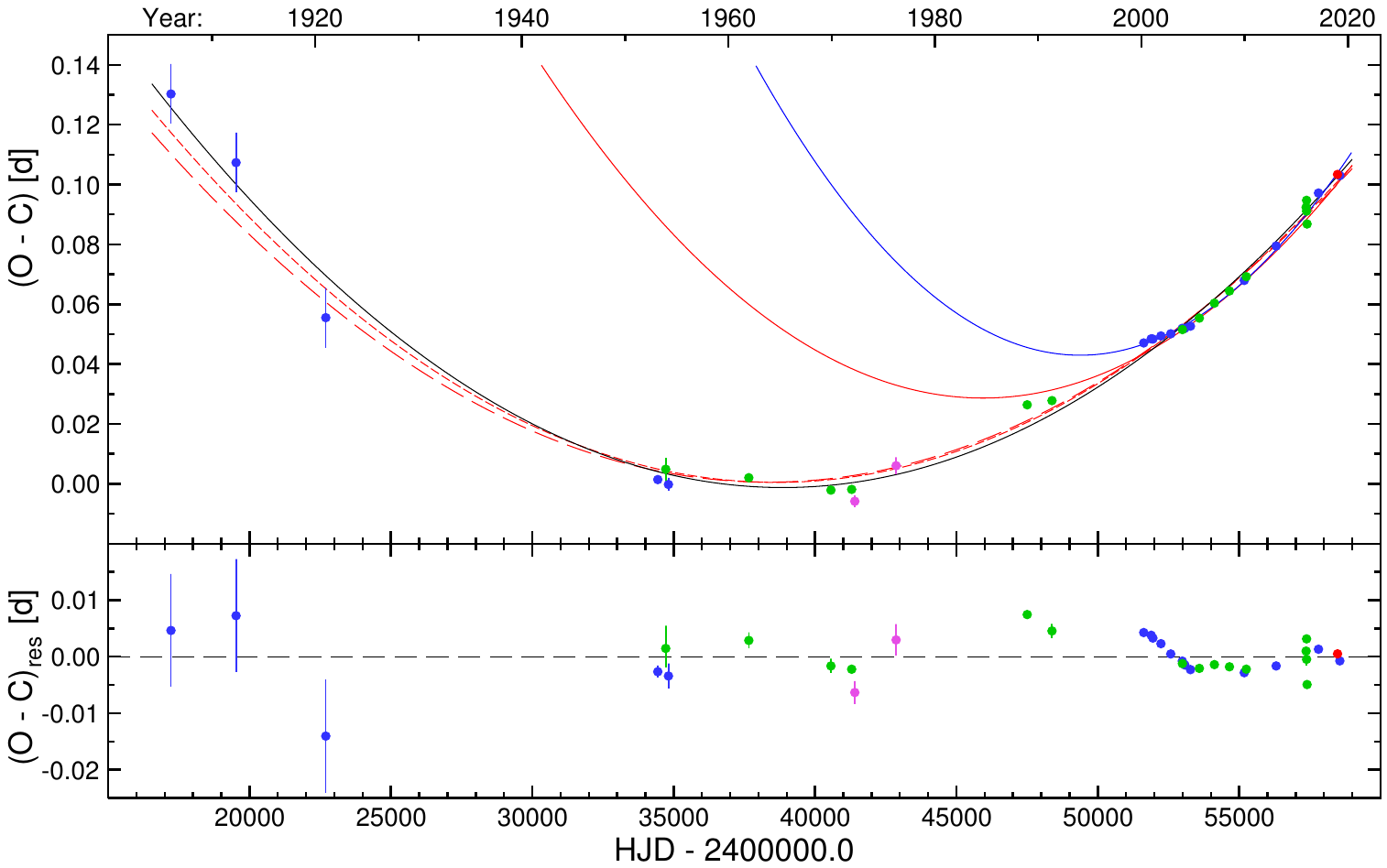}
\caption{Top: {\oc} diagram for the light (green dots) and radial velocity (blue dots) times of maximum given in Tables \ref{tab:tmax-rvel} and \ref{tab:tmax-phot}. A constant shift of $+$0.070~d was applied to the times of maximum radial velocity. Two violet dots correspond to $T_{\rm max}$ derived from TD-1A and ANS UV observations, {while the red dot corresponds to $T_{\rm max}$ derived from the TESS data. These three} values of  O\,--\,C are not considered in the fits. Four of the parabolas are fits with different weighting schemes: weights proportional to $\sigma^{-2}$ (red continuous line), $\sigma^{-1}$ (red long-dashed line), $\sqrt{\sigma}$ (red short-dashed line) and with equal weights (black line). The blue line shows the fit to the recent {(2000\,--\,2019)} RV data only ({$\dot{P}=$ 0.97\,$\pm$\,0.13~s/cen} with equal weights). Bottom: residuals from the fit with equal weights. }
\label{fig:o-c}
\end{figure*}

The radial velocities of {\xic} were derived from the optical spectra and the analysis rarely included hydrogen lines. Therefore, systematic effects related to the velocity gradient  in the atmosphere and non-adiabaticity, resulting in the phase lag between hydrogen and other lines, called van Hoof effect \citep{1953PASP...65..158V,1991A&A...252..245M}, is probably negligible in the case of {\xic}. On the other hand, the lack of phase lag cannot be assumed for photometric data, especially for a radially pulsating star with large amplitude like {\xic}. There are two multicolour data sets, which potentially enable to check if the phase lags are observed for {\xic}. The first one is the UV TD-1A photometry of {\xic} in four bands published by \cite{1980AAS...39..301B}, the other one is the seven-band Geneva photometry of \cite{1992AAS...96..207H}\footnote{Kindly provided by Gerald Handler.}. In both data sets, the observations in all passbands are simultaneous. The effective wavelengths of these 11 passbands range between 151 and 570~nm. The Geneva photometry covers a wider wavelength range and is of better quality than the very scarce (8 data points only) TD-1A time-series. Due to the scarcity, the latter data set cannot be used to conclusively discuss the phase lags in the UV. The situation is much better for the Geneva data. The derived times of maximum light obtained from the Geneva data are plotted against the effective wavelengths of the passbands in Fig.\,\ref{fig:pl}. There is a clear dependence of the time of maximum light on $\lambda_{\rm eff}$ --- the longer the wavelength, the later the time of maximum occurs. A least-squares fit gives the rate of the time of maximum lag $A_{\rm Vis}= (+$3.99\,$\pm$\,0.33)\,$\times$\,10$^{-5}$~d\,(nm)$^{-1}$. The derived value of $A_{\rm Vis}$ translates into a difference in the time of maximum light equal to 0.0090~d $=$ 13~min or 7 per cent of the pulsation period in the full range of the effective wavelengths covered by the Geneva filters (348\,--\,570~nm). Because simultaneous UV and visual data for {\xic} do not exist, the phase-lag corrections cannot be applied to the UV data. Consequently, we do not use the UV data in the fits shown in Fig.\,\ref{fig:o-c} (although they are shown for reference). {Similarly, the phase lag is not extrapolated and the time of maximum not corrected for the TESS passband.} The effect of phase (or times-of-maximum) lag has to be taken into account to properly use photometric data obtained in different passbands in the {\oc} diagram. These corrections to the times of maximum light in the visual domain, $C_{\rm Vis} \text{[d]} = A_{\rm Vis}(500 - \lambda_{\rm eff})$, where $\lambda_{\rm eff}$ is in nm, were added to the observed times of maximum light, $T_{\rm max}^{\rm obs}$:
\begin{equation}\label{eq-corr}
T_{\rm max}^{\rm corr} = T_{\rm max}^{\rm obs} + C_{\rm Vis}
\end {equation}
before the {\oc} values were calculated. Both $T_{\rm max}^{\rm obs}$ and $C_{\rm Vis}$ are reported in Table\,\ref{tab:tmax-phot}. The values of $T_{\rm max}^{\rm corr}$ were subsequently used to calculate the values of {\oc} according to the ephemeris, taken from \cite{1973MNRAS.162...25S}:
\begin{equation}
T_{\rm max} = \mbox{HJD} 2441296.0514 + 0.2095755 \times E,
\end{equation}
where $E$ is the number of periods elapsed from the initial epoch. The same ephemeris was used for radial velocity times of maximum (Table \ref{tab:tmax-rvel}).

Whenever applicable and possible, the uncertainties of $T_{\rm max}$ were derived by means of the bootstrap method. For samples with small numbers of data points (ANS photometry and the oldest radial velocities), the uncertainties were inferred from the least-squares variance and multiplied by 4. This number was taken from the comparison of uncertainties derived from least-squares and bootstrapping errors for slightly more numerous but still small samples of data. For the data published by \cite{1962ZA.....56..141V}, the published values of $T_{\rm max}$ were averaged after transferring to the same (mean) epoch, and the uncertainty was estimated as a standard deviation. Finally, the published uncertainties of $T_{\rm max}$ (when no data were available) were multiplied by two as a conservative choice, because usually no details of their derivation were given.

%\comment{GAW: Andrzej, can you confirm that the red continous line is really at $\sigma^{-2}$-weighted fit to all of the data points? I'm surprised at how discrepant it is. APi: Confirmed. This is the last RV point which makes the difference. The old data, due to strong weighting, become insignificant and we get what you can see.}s

\begin{figure}
\centering
\includegraphics[width=\columnwidth]{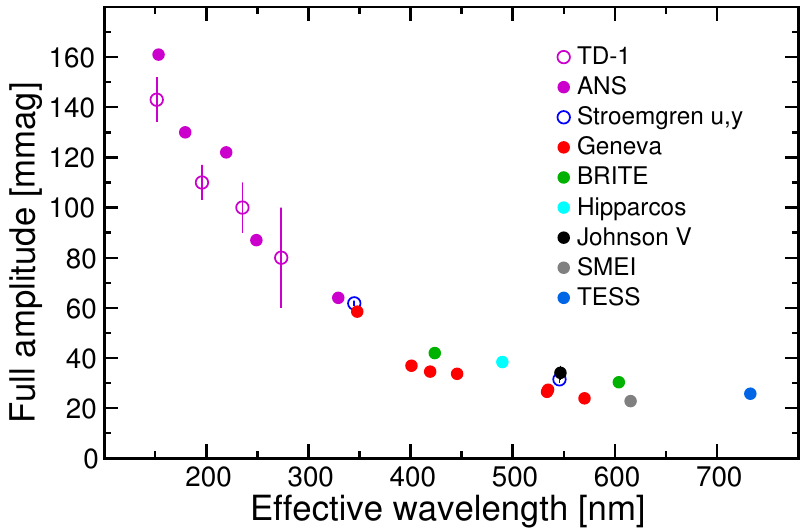}
\caption{{Full amplitudes} of the radial mode of {\xic}. The labels indicate either space mission (TD-1, ANS, Hipparcos, BRITE, SMEI, {TESS}) or photometric system. The observations can be identified in Table \ref{tab:tmax-phot}.}
\label{fig:ampl}
\end{figure}

%\comment{This paragraph has been moved here, as Fig.\,\ref{fig:ampl} is referred to earlier in the text.}
As a by-product of the procedure of the determination of times of maximum light, we obtained amplitudes of the radial mode. The amplitudes are shown in Fig.\,\ref{fig:ampl} as a function of $\lambda_{\rm eff}$. A strong increase of amplitude towards short wavelengths, typical for radial modes in $\beta$~Cep stars, can be seen. In addition, a small amplitude change could have taken place in {\xic}. For example, the amplitudes in the two BRITE bands are about 25\% larger than those derived from the Geneva data. The two data sets are separated by almost three decades, however.

%\comment{Should we now show the unrealistic fit with weights $\propto\sigma^{-2}$?}

\subsection{Correction for the phase lag between light and radial velocity changes}\label{RVcorr}
The {lag} between light and radial velocity data was derived by a trial and error procedure using photometry from SMEI and BRITE and the recent radial velocity measurements. The phase lag is equal to $+$0.070\,{$\pm$\,0.001~d}, corresponding to 0.334\,{$\pm$\,0.005} in phase. This number is very different from a typical value of 1/4, corresponding to the maximum light at the epoch of minimum radius, and observed in other $\beta$~Cep stars. Even for the two high-amplitude $\beta$~Cep stars, BW~Vul and $\sigma$~Sco, the phase shifts amount to 0.249 \citep{1993A&A...274..269P} and 0.265 \citep{1992A&A...261..203P}, respectively; that is, much less than in {\xic}. 

{Nonlinear pulsational calculations indicate that BW Vul is almost certainly a fundamental radial pulsator \citep{1994IAUS..162...19M}. In this case, the observed standstill in the light curve is caused by an emerging shock wave which originates at the bottom of the He\,{\sc ii} ionization zone. The first overtone mode is stable.}

%\comment{I have some plots prepared by H.Cugier to explain the phase lag. It seems it can be explained by the models. Maybe it would be useful to add a small paragraph on this?}
\subsection{Seismic inference from phase lags}\label{seisminf}
The phase lag between light and radial-velocity curves as well as the dependence of phase of the maximum light on wavelength (Fig.\,\ref{fig:pl}) can possibly be used to constrain stellar parameters or verify mode identification for the dominant mode. The 4.77~d$^{\rm -1}$ frequency has  already been identified as a radial mode by \cite{1994A&AS..105..447H}, \cite{1994A&A...291..143C}, and \cite{2006CoAst.147..109S} using both photometry and spectroscopy. The strong dependence of amplitude on wavelength (Fig.\,\ref{fig:ampl}) provides a clear indication that this is the case. However, it was not clear if the mode was fundamental or an overtone {although \cite{2017MNRAS.471.2286S} showed that the stellar parameters they derived are consistent with the fundamental mode.} %\comment{We claim in the introduction that the mode is fundamental, but I could not found the source of this information (I mean who found it is fundamental). If this was not shown earlier, the word 'fundamental' should be removed from the first paragraph of the paper. This also means that identifying the mode as fundamental is the result of this paper. We need to check this, however.}

We therefore checked if seismic models are able to reproduce the two key observed characteristics: the large phase lag between light and radial-velocity curves, and the phase dependence on wavelength. For this purpose, we calculated a grid of models for stars with stellar parameters close to those provided by \cite{2017MNRAS.471.2286S}. The models were calculated in the same way as described by \cite{2014A&A...565A..76C}. We used {OPAL} opacities, hydrogen mass abundances {between $X = 0.6$ and 0.8 and two metallicities, $Z = 0.0134$ and 0.0168.
%Moreover, we included core overshooting assuming overshooting parameter $\alpha_{\rm{ov}}$ in the range between 0 and 0.6.
} 
%We did not include core overshooting, but mixing of elements was taken into account within the boundary of the convective core determined by Schwarzschild criterion. 
Pseudo-rotating, spherically symmetric models (following \citealt{1998A&A...334..911S}) were built assuming rigid rotation with constant total angular momentum during the main sequence evolution. The models were also used to calculate amplitude ratios and phase differences for the Str\"{o}mgren, Geneva and BRITE photometric systems following the procedure presented by \cite{1994A&A...291..143C}. {We used LTE models calculated by \cite{2003IAUS..210P.A20C} (hereafter referred to as CK models) and non-LTE models calculated by \cite{2007ApJS..169...83L} using the TLUSTY code \citep{1988CoPhC..52..103H,2011ascl.soft09021H} for microturbulent velocity of $\xi = 2$\,{\kms} (BG models). The latter grid of non-LTE models has been extended for $\log g > 3.0$ assuming $\xi = 10$\,{\kms} following the procedure described by \cite{2012A&A...547A..42C}.}

{The first conclusion that can be drawn from these calculations is that} the frequency of the mode for models with masses between 14.0 and 14.5\,M$_\odot$ and $T_{\rm eff}=27$\,kK, consistent with parameters of {\xic} provided by \cite{2017MNRAS.471.2286S}, can be reproduced only if the radial mode is fundamental. A first overtone {can be excluded because: (i) It} would require much smaller $T_{\rm eff}\approx 24.5$\,kK, which is neither compatible with the spectrum nor the colours of {\xic}.  {(ii) It is stable in models. (iii) The theoretical phase lag between light and RV curve is smaller than 0.25 for all models in the considered range of masses, in contrast with the observed value. (iv) As already concluded by \cite{2017MNRAS.471.2286S}, the stellar parameters they derived are consistent with thes pulsation constant corresponding to the fundamental radial mode. For an overtone, the mass and luminosity inferred from the pulsation constant would be much too high. The second and higher overtones can be excluded for the same reasons. Therefore, we conclude that the observed variation corresponds to the fundamental radial mode.}

{As can be seen in Fig.\,\ref{fig:pl}, the wavelength dependence of the photometric phase lags is well reproduced by non-rotating BG models with $M=14$\,$M_\odot$, $T_{\rm eff} = 27$\,kK, $X=0.7042$, $Z=0.0162$, and $\log g = 3.74$ provided that it is assumed that the mode is radial fundamental. The same models predict phase lags between light and RV curves in the range 0.38\,--\,0.39, slightly too high in comparison with the observed value. For the non-rotating CK models with $X=0.7374$ and solar metallicity ($Z=0.0134$), the phase lag equals to 0.35\,--\,0.37. Although still slightly higher than the observed value of 0.334\,$\pm$\,0.005, this value can be regarded as fairly consistent with the observations given the non-sinusoidality of the light curve, which is not reproduced by the models.}
    
\subsection{The resulting {\oc} diagram}
The {\oc} diagram, which uses both photometry and spectroscopy, is shown in Fig.\,\ref{fig:o-c}. Given the uncertainties of the times of maximum light and radial velocity, the changes of pulsation period cannot be perfectly approximated by a simple parabola corresponding to $\dot{P} =$~const, although such a model is relatively good as a first approximation. The residuals shown in the lower panel of Fig.\,\ref{fig:o-c} are much larger than the associated uncertainties (if they cannot be seen, they are smaller than the size of the symbols). Due to the inadequacy of the fitted model, a typical weighting scheme (weights $\propto \sigma^{-2}$) is not the best choice in this case (a large range of uncertainties) leading to large residuals. In total, four different weighting schemes were tried and the results are shown in Fig.~\ref{fig:o-c}:
\begin{enumerate}
\item weights $\propto\sigma^{-2}$, $\dot{P}= 0.603\pm 0.090$~s/cen.
\item weights $\propto\sigma^{-1}$, $\dot{P}=$ 0.325\,$\pm$\,0.024~s/cen. 
\item weights $\propto\sigma^{-1/2}$, $\dot{P}=$ 0.338\,$\pm$\,0.012~s/cen. 
\item equal weights, $\dot{P}=$ 0.358\,$\pm$\,0.008~s/cen. 
\end{enumerate}

%\comment{[Question from Tony for APi: I understand gaussian weights of $1/\sigma^2$ and equal wts. But how do you justify $1/\sigma$ or $1/\sqrt{\sigma}$?]} 

The values are consistent with 0.37 $\pm$ 0.05 s/cen reported by \citet{1999NewAR..43..455J} and derived {by \cite{1992PhDT-Pigulski}}.

As mentioned above, the residuals from the best-fit parabola exhibit scatter that is larger than the uncertainties. In particular, the dense photometric and spectroscopic sampling since the year $\sim 2000$ is clearly incompatible with the long-term trend, and suggests more complex and rapid period variations. This is confirmed when we attempt to phase these recent measurements using $\dot P=0.3$~s/cen: the measurements are not coherently phased, with clear phase offsets between the datasets. As demonstrated by \citet{2017MNRAS.471.2286S} and \citet{2018pas8.conf..154B}, a much larger $\dot P\sim 0.9$~s/cen rate of period change is needed in order to reconcile them. A period search of the residuals shown in the bottom panel of Fig.~\ref{fig:o-c} yields weak evidence for a period around 40\,yr.

\section{Interpretation}
Examination of the \oc\ diagram shows that there are two phenomena to be explained: the longer-term increase of the pulsation period at a rate of $\sim$0.3~s/cen, and the more rapid variations detected by \citet{2017MNRAS.471.2286S}. Because the more rapid changes are diagnosed principally by the modern data, it is unclear if they are a recent phenomenon, or if they existed all along and were only revealed by the recent high-precision observations (although the significant scatter of much of the modern data on the \oc\ diagram suggests the latter). 

In this Section we examine the potential contributions of various phenomena (binarity, stellar evolution, additional (undetected) pulsation modes, and stellar rotation/magnetism) to the long- and short-term evolution of the (apparent) pulsation period.

%\textcolor{comment}{[This could be interpreted in the terms of the light-time effect in a binary system. The picture is not clear, however, due to the gaps in data and this will be really difficult to prove (or disprove).] \Matt{(Well, my binarity analysis wasn't really checking for a companion from this standpoint. All I could do is rule out a star bright enough to cause the Halpha emission. I guess a dimmer companion might be able to cause the effect we're looking at. We could look for this using RV residuals, using the ephemeris in this paper. I suspect there will be a very low-frequency signal; if so, the question is whether the RV amplitude and period could be consistent with a companion capable of causing this effect.)}}

%The apparent evolution of the period of {\xic} may have several origins. 

\subsection{Binarity}

The light-time effect in a binary system is well known to result in apparent changes of period of orbiting pulsating stars \citep[e.g.][]{1992A&A...261..203P}. The change in pulsation period is given by: 
\begin{equation}
\Delta P=P\, \Delta V_{\rm r}/c
\end{equation}
where $\Delta P$ is the predicted change of pulsation period $P$ due to light-time effects associated with a radial velocity variation $\Delta V_{\rm r}$, and $c$ is the speed of light. 

\citet{2017MNRAS.471.2286S} searched using PIONIER in particular for a Be star companion to \xic. They were able to rule out any companion brighter than 1.7\% of \xic's flux (in the $HK$ bands) beyond 40 AU, with a similar upper limit derived from the standard deviation of the RVs, within about 40 AU. {\xic} is reported in the {Washington Double Star Catalogue} (WDS) to have a companion ($V=14$~mag) located at 28\arcsec\ from the primary. At the Hipparcos distance of {\xic} (424 pc\footnote{Gaia DR2 gives $\pi=4.984\pm 0.346$~mas. This is completely incompatible with the Hipparcos parallax used by \citet{2017MNRAS.471.2286S}. It is also less precise. At the corresponding distance of about $200 \pm 14$ pc, the star would have a luminosity of $\log{L} = 3.83 \pm 0.06$, which at its effective temperature would place it on the Zero-Age Main Sequence, an age at which it would be unlikely to be a $\beta$ Cep pulsator. {As Gaia parallaxes of bright stars ($V\ltsim 6$) should, for the time being, be considered with caution \citep{2018A&A...616A...2L}, we adopt the Hipparcos parallax.}}), one arcsec is {424 AU, so this separation would correspond to nearly 12000 AU}. The flux difference is nearly 10 mag in the $V$ band, so we estimate that the companion (assuming it is located at the same distance) would be a K dwarf with a mass of $\sim$0.7\,$M_\odot$. The resultant periods are far too long to explain the observed \oc\ diagram.
% From WDS
%06319-2325SEE  68AB    1897 1999    4 147 147  24.8  26.5  4.33 14.0  B0.5IV    -003+006          -23 3991 N    063151.37-232506.4
%06319-2325SEE  68AC    1897 1999    5 303 304  28.9  27.6  4.33 14.2  B0.5IV    -003+006 -009+017          N    063151.37-232506.4
\begin{figure}
  \centering
    \includegraphics[width=0.45\textwidth]{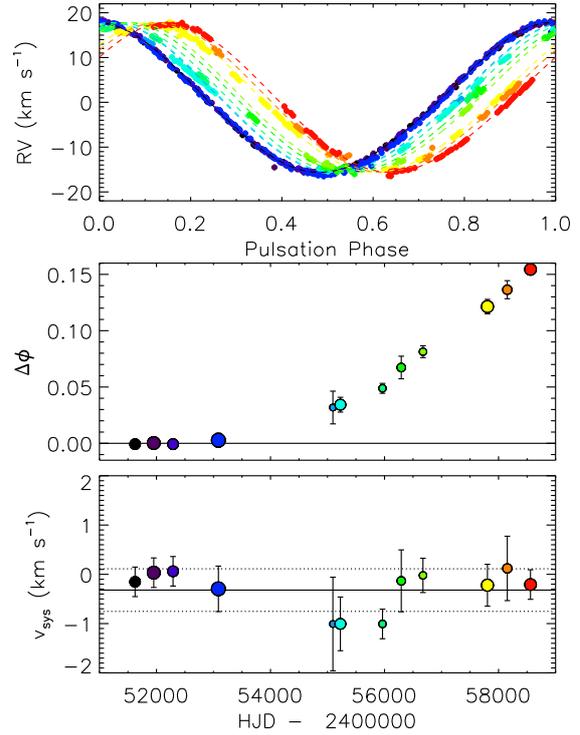}
    \caption{{\em Top:} CORALIE and ESPaDOnS RVs phased with the pulsation period determined from the first epoch of CORALIE data. Colours indicate {time bins}. Dashed lines show 3$^{\rm rd}$-order harmonic fits to the CORALIE data, shifted to minimize the standard deviation of the residuals. {\em Middle:} phase shift $\Delta\phi$ of the RV curve in each epoch. {\em Bottom:} systemic velocity $v_{\rm sys}$ determined from the mean residual RV after subtraction of the phase-shifted curves in the top panel. {The solid line shows the mean $v_{\rm sys}$, the dotted lines the standard deviations.} %The dashed curve relates to the most conservative scenario for the light-time hypothesis (see text). {In the middle and bottom panels, symbol size is proportional to the log of the number of points in each bin.}
    }
    \label{rv_zero}
\end{figure}

To attempt to measure the systemic radial velocity $v_{\rm sys}$ of \xic, we fit the RVs from the first year of CORALIE data described by \citet{2017MNRAS.471.2286S} with a 3$^{\rm rd}$-harmonic fit, shifted this curve in phase in each successive {two-year bin} in order to minimize the standard deviation of the residuals, and then calculated the mean residual RV after subtraction of the curve. The results of this exercise are shown in Fig.\,\ref{rv_zero}. As can be seen in the bottom panel, the RV curve is consistent with no change in $v_{\rm sys}$ to within about {$\pm 0.4$} \kms~over the span of observations. {If the 0.9 s/cen period change is due to orbital motion, then it should have corresponded to a change in RV of about 3 \kms~over the approximately 20-year span of the RV observations. Since this would have been easily detected, binarity can be ruled out as the source of this period change.}

%{The 0.9 s/cen period change seems to be a faster variation superimposed on the longer-term 0.3 s/cen period change, with a characteristic time scale on the order of 40 years.} This rules out the light-time effect as the source of the 0.9 s/cen period change, as (assuming a periodicity of 40 years) a 15 \kms~RV shift during the $\sim$20 years spanned by these data would have been easily detected. 

If the 0.3 s/cen period change is due to binarity, on the other hand, we would expect a maximum RV shift of about {1} \kms~over the {20 years} covered by these data. {Since this is comparable to the standard deviation of $v_{\rm sys}$, orbital motion cannot be excluded in this case. However, there is no positive evidence for a change in $v_{\rm sys}$, and as will be shown below in Sect.\,\ref{stelev} there is good reason to believe that the 0.3 s/cen period change is due to stellar evolution.}

%Assuming the orbit to be approximately circular, the orbital RV curve should be sinusoidal. In this case, if we were particularly unlucky and our RV measurements correspond to an extremum, the rate of change of RV would be very low and the resulting RV shift could, indeed, remain undetectable within these data. This is illustrated in the bottom panel Fig.\,\ref{rv_zero}, where the dashed line shows a sinusoid with an amplitude of 5 \kms, shifted in velocity to account for an arbitrary systemic velocity.

%While it cannot be excluded that the 0.3 s/cen period change is due to binarity, at almost any orbital phase other than the extremum the RV shift due to the companion would have been detected; furthermore, there is no positive evidence for an RV shift, and a
\begin{figure}
  \centering
 % \begin{minipage}{\textwidth}
    \includegraphics[width=0.5\textwidth]{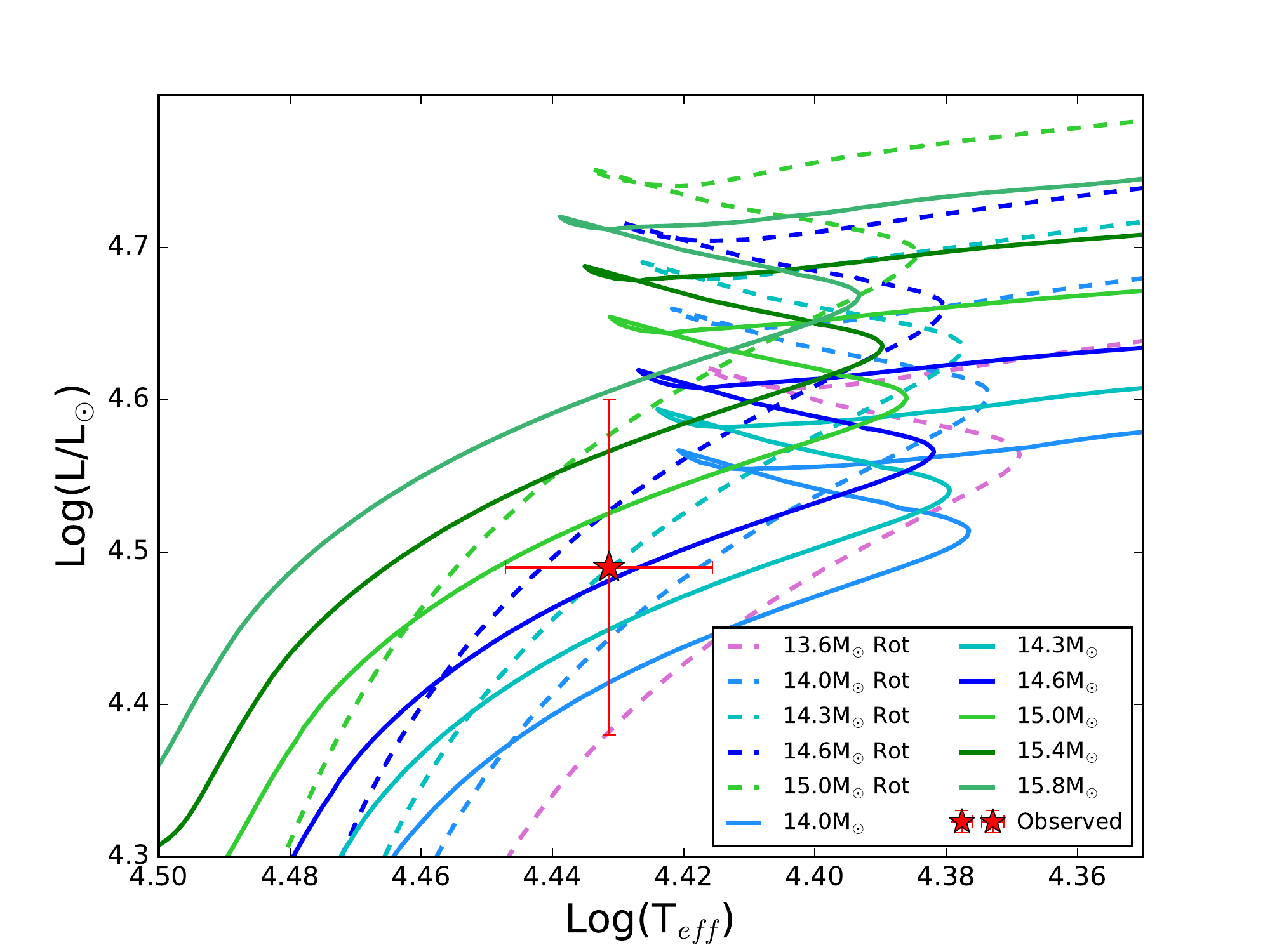}
    \includegraphics[width=0.5\textwidth]{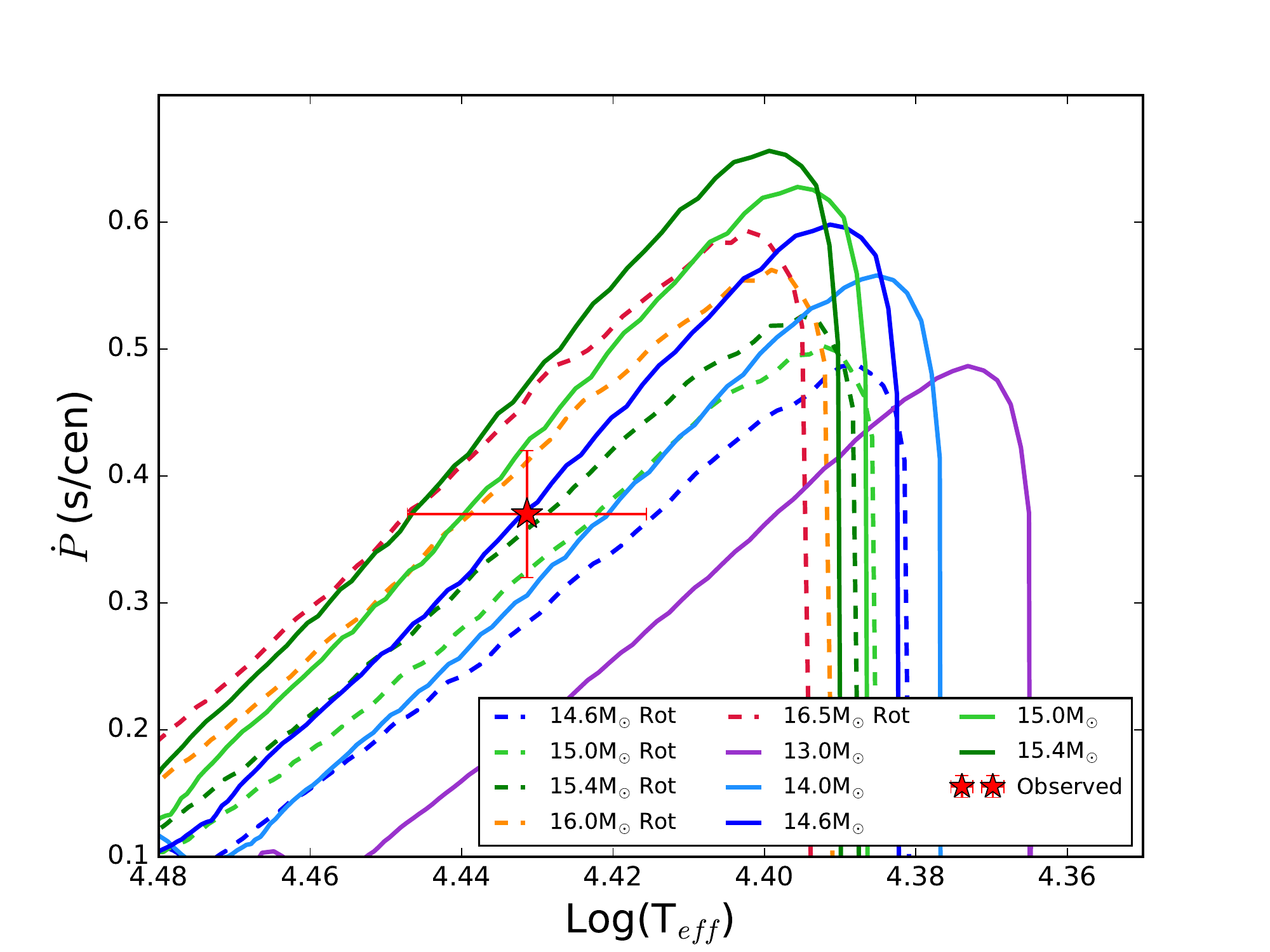}
\caption{{\em Upper panel:}\ Theoretical HR diagram showing evolutionary models ignoring and including the effects of rotation, calculated by \protect{\citet{2012AA...537A.146E}}. {\em Lower panel:}\ $\dot{P}-T_{\rm eff}$ plane showing the same evolutionary models {above}, calculated according to Eq.(\ref{pdot_evol}) using the evolutionary models. The positions of {\xic} according to the physical parameters inferred by \protect{\citet{2017MNRAS.471.2286S}} and the rate of period change determined here are shown in red.}
    \label{HRD}
  %\end{minipage}
\end{figure}

\subsection{Stellar evolution}\label{stelev}
Since the period is growing, this is qualitatively consistent with the increasing radius of the star as expected due to stellar evolution on the main sequence \citep[e.g.][]{2015A&A...584A..58N}. As reported by those authors, the fractional rate of change of the pulsation period of a radially pulsating star due to evolving mass $M$ and radius $R$ on evolutionary timescales can be computed according to:
\begin{equation}
\frac{\dot{P}}{P} = -\frac{1}{2}\frac{\dot{M}}{M}+\frac{3}{2}\frac{\dot{R}}{R}.\label{pdot_evol}
\end{equation}

We have exploited the evolutionary model calculations of \citet{2012AA...537A.146E} to predict the variation of $\dot{P}$ according to Eq.\,(\ref{pdot_evol}). Given that {\xic} is a (relatively) cool upper-main sequence star, its mass-loss rate is expected to be low; hence we have assumed $\dot{M}=0$ in Eq.\,(\ref{pdot_evol}).

%The physical parameters of {\xic} determined by \citet{2017MNRAS.471.2286S} place the star on the second half of the main sequence. 

In Fig.~\ref{HRD} we show the star's position on the Hertzsprung-Russell (HR) diagram and on the $\dot{P}$ vs. $T_{\rm eff}$ diagram, using the physical parameters of \citet{2017MNRAS.471.2286S} and $\dot P=0.32\pm 0.05$~s/cen, and models both including the effects of rotation ($v_{\rm rot}=200$~km/s) and ignoring those effects \citep{2012AA...537A.146E} (These tracks bracket potential evolutionary histories of \xic; although the star is known to be a very slow rotator today, the rotational history of the star, and hence the appropriate evolutionary tracks, are unknown.).  For the non-rotating tracks, we derive a best-fitting mass of 14.6~\msun\ and age of 9.2~Myr. For the rotating tracks, we derive a best-fitting mass of 14.4~\msun\ and age of 11.1~Myr. The masses are formally consistent with that derived by \citet{2017MNRAS.471.2286S}. The position on the $\dot P-T_{\rm eff}$ is formally consistent with both sets of models. We note that the more recent $\dot P\sim 0.9$~s/cen rate of period change does not agree with the models and derived $T_{\rm eff}/\log L$.

%\comment{[Hilding comments: "One thing that stuck out to me, though probably won't to a referee or be significant is the plot of period change and effective temperature in Fig 5.  I assume the evolution model period change is computed using Eq. 6, hence what is the period for the models.  Is the period assumed to be the observed value or is it computed separately? If it is assumed to be the observed value the rates of period change in the models could be a bit different.  When I do a quick check using the period-mean density relation, the range of Teff and Log L for observations in Fig 5 I find that the pulsation period will vary by about a factor of two.  I would suggest replacing the second plot (P-dot vs Teff) with  the relative rate of period change so that the model rates are based on evolutionary considerations only (P-dot/P vs Teff)." Gregg replied: ``Hilding, yes, the P-dot is assumed constant. Are you suggesting that the current P-dot axis be scaled according to the period-mean density relation?"]} \modif{Gregg replies: I still don't understand the motivation for doing this, which suggests to me that I've misunderstood what is being asked. How does dividing the y-axis by a constant change anything?}

%, although in somewhat better agreement with the non-rotating models.

\subsection{Undetected pulsation modes}
To interpret 25-year period variations in the O-C diagram of the $\beta$~Cep star BW~Vul, \citet{1984PASP...96..657O} submits that a second pulsation mode close in frequency to the primary model could result in the apparent period variation: ``An alternate interpretation of the behaviour of BW~Vul is in terms of two pulsation modes which are so close to the same period that they don't 'beat' in the normal sense. In this case, the smaller-amplitude mode would have its peak first on the rise up to the peak of the large- amplitude mode, thus making maximum brightness earlier than usual." In the case of BW~Vul, the primary pulsation period is very similar to that of \xic, but in our case no obvious O-C periodicity is observed. However, if we assume that the $\sim$40~yr timescale of O-C variability is a result of such a model, then according to \citet{1984PASP...96..657O} the second mode would differ from the known radial pulsation period by about 0.25\,s. In any case, as mentioned by \citet{1984PASP...96..657O}, such unresolved beating should also result in amplitude changes in accordance with apparent period changes. This is not seen for for BW~Vul nor \xic.

%There is evidence (see below) that the spectrum of \xic\ is modulated according to the rotational period due to multiple phenomena.

%Magnetosphere is being modulated. \Matt{(I suppose the magnetosphere could lead to some degree of phase shifting. It should also produce a variation in the total brightness, though. The magnitude of this effect could probably be checked via ADM. Given how weak the emission is, however, I'm skeptical that it would be very significant. It's also not clear to me that this would affect the RVs; weak Halpha should mean undetectable metallic emission.  Maybe spots? But for such a hot star, that would be unexpected. That said, there are long-term variations in the strength of the metal lines. I have no idea what's causing them. I'll send plots so you can see what I mean.)}

\subsection{Stellar rotation \& magnetism}
As mentioned earlier, the O-C residuals show weak evidence for a periodic behaviour with $P \sim$ 40 years. Such a timescale could be compatible with rotation of \xic. Magnetic early-type stars typically exhibit line profile variability coherent with the rotation period. This can be a consequence of Zeeman splitting, surface chemical abundance peculiarities, or magnetospheric emission. In principle these effects can affect radial velocity measurements. The period change apparent in the more recent data might be oscillatory, and possibly with a decadal timescale; since the rotation period of {\xic} is extremely long \citep[at least 30 years;][]{2017MNRAS.471.2286S,2018MNRAS.478L..39S}, it is natural to wonder if some form of rotationally modulated variation might be influencing the radial velocity or light variation of the star. 
\begin{figure*}
  \centering
	\begin{tabular}{ccc}
    \includegraphics[trim = 50 50 0 0, width=0.333\textwidth]{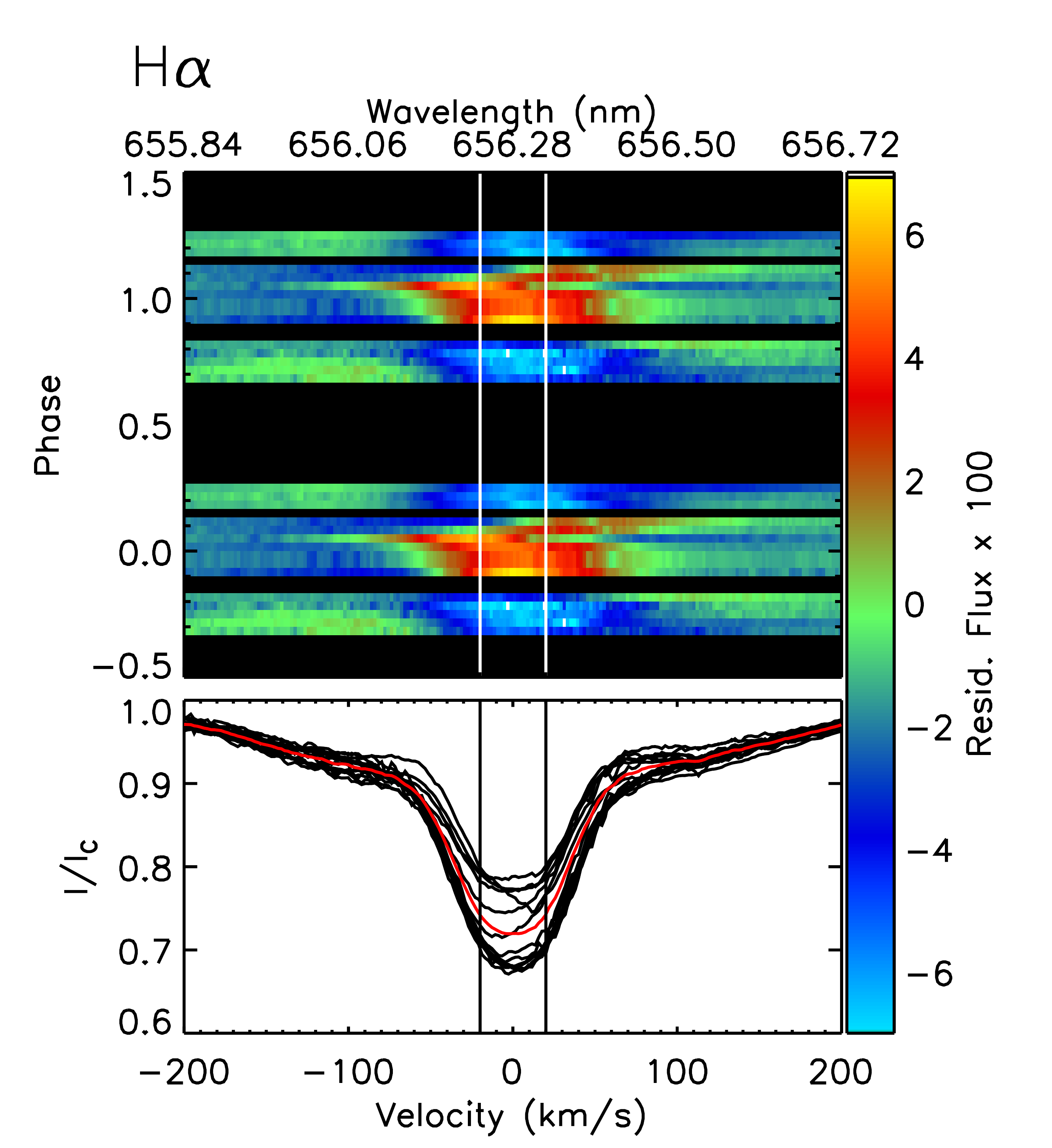} & 
    \includegraphics[trim = 50 50 0 0, width=0.333\textwidth]{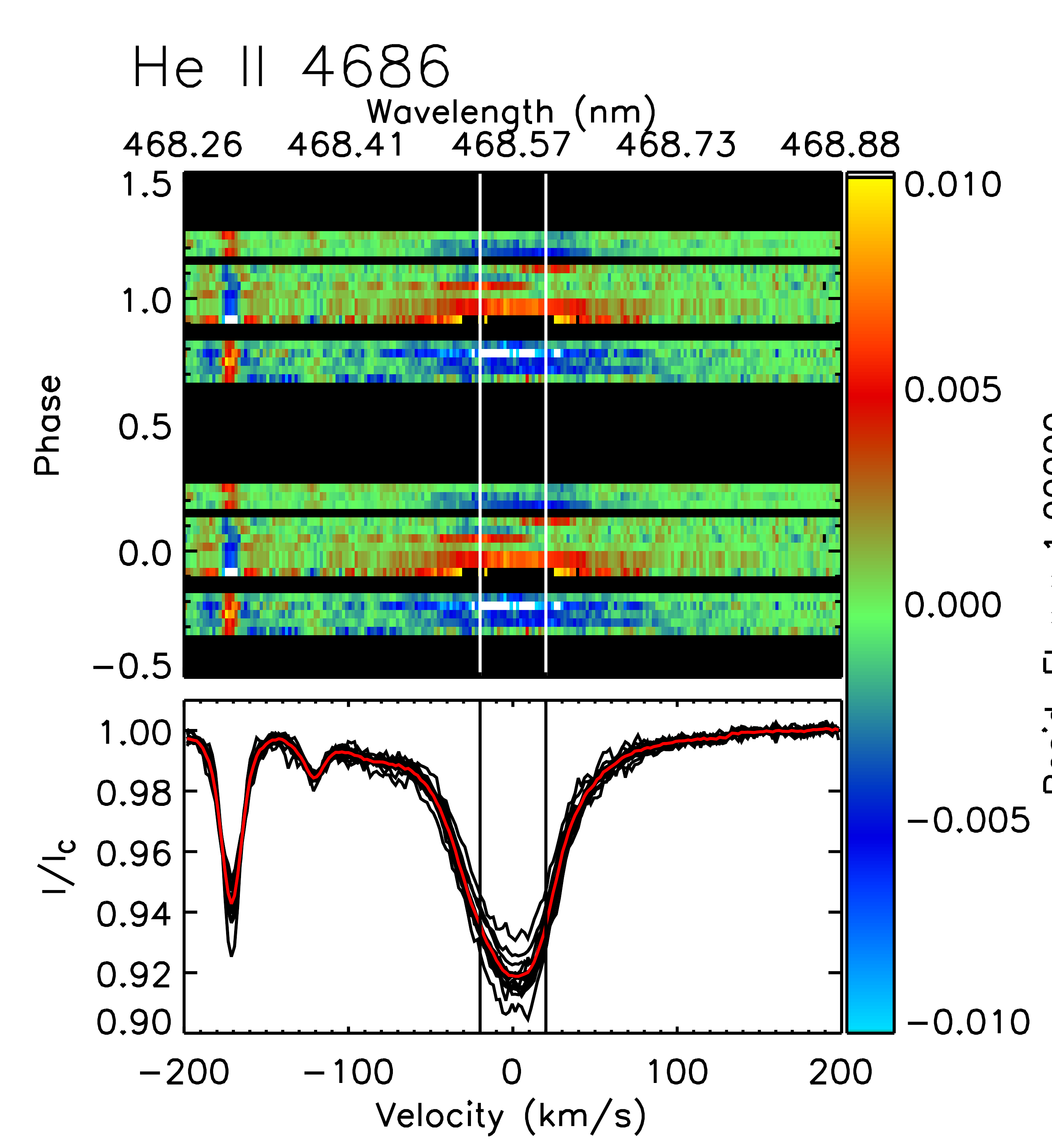} & 
\includegraphics[trim = 50 50 0 0, width=0.333\textwidth]{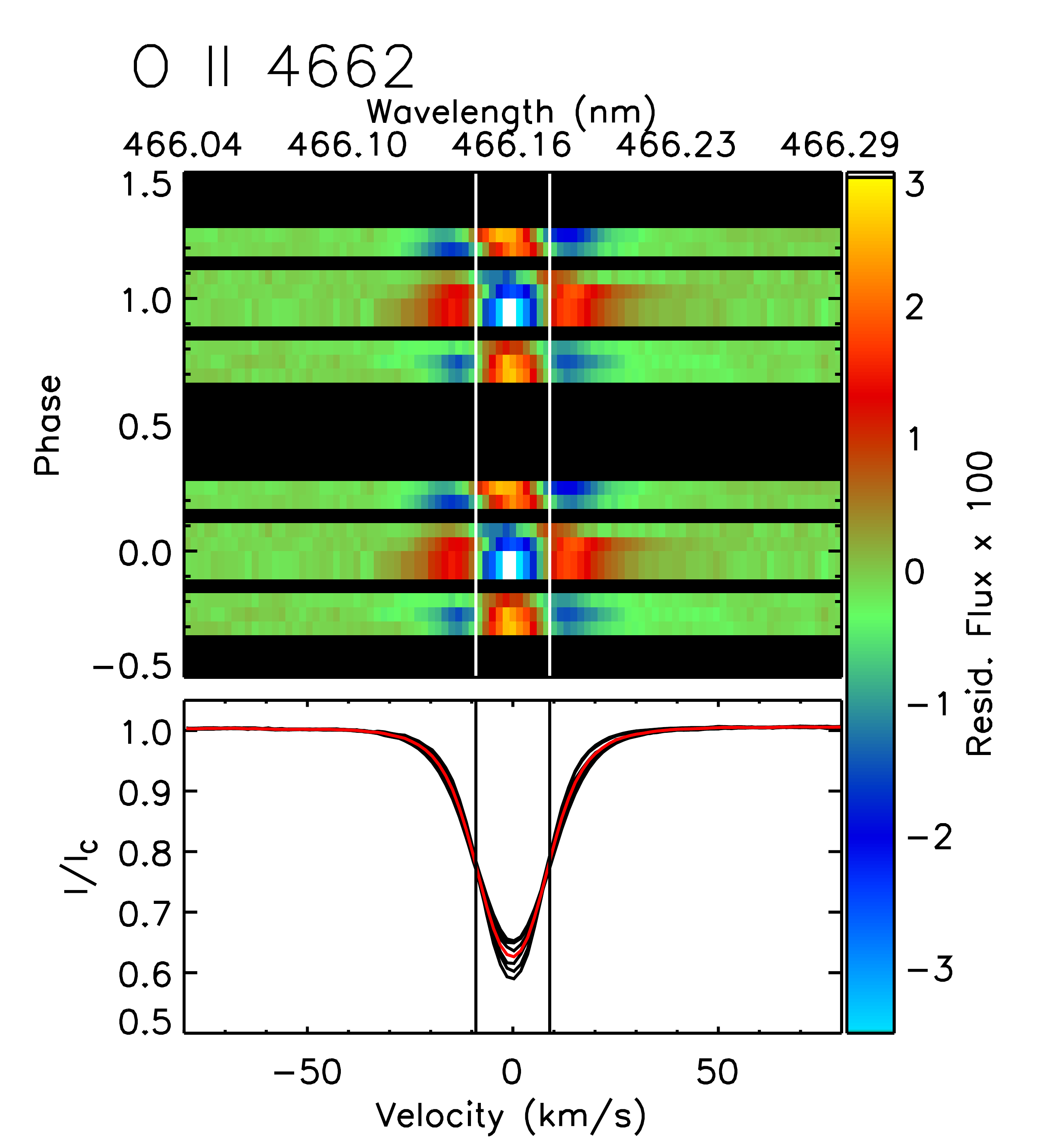} \\
\includegraphics[trim = 50 50 0 0, width=0.333\textwidth]{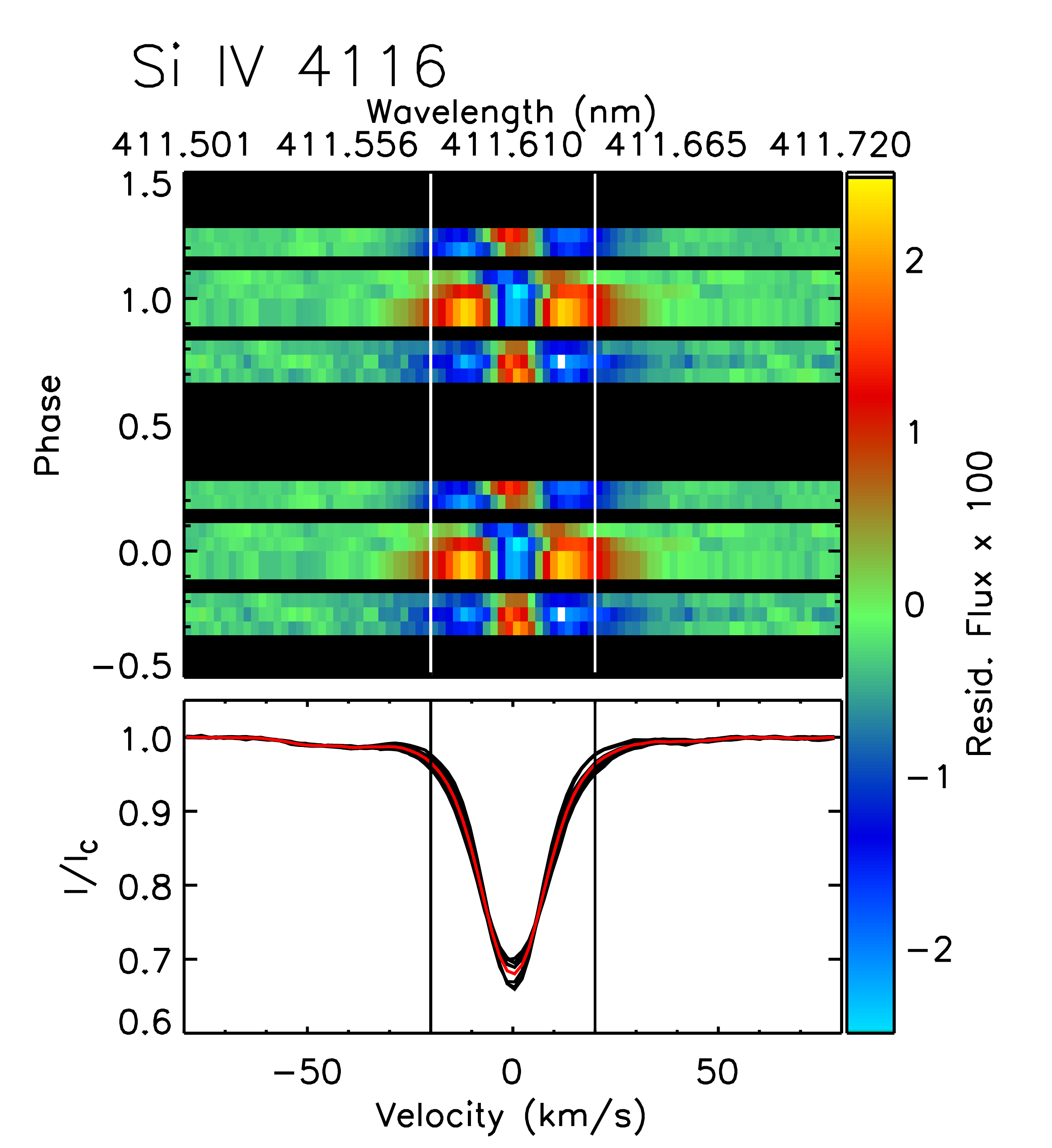} & 
\includegraphics[trim = 50 50 0 0, width=0.333\textwidth]{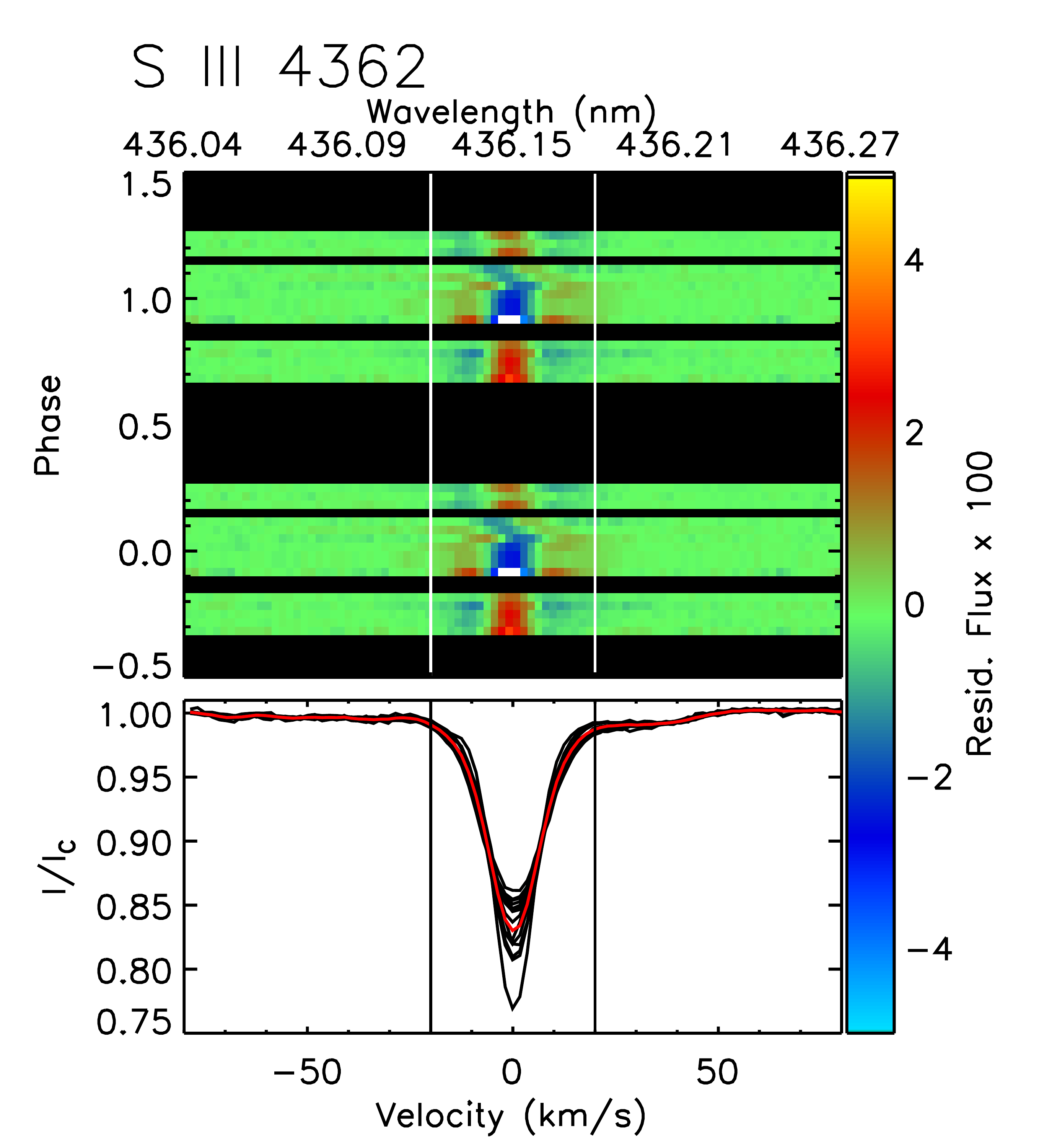} &
\includegraphics[trim = 50 50 0 0, width=0.333\textwidth]{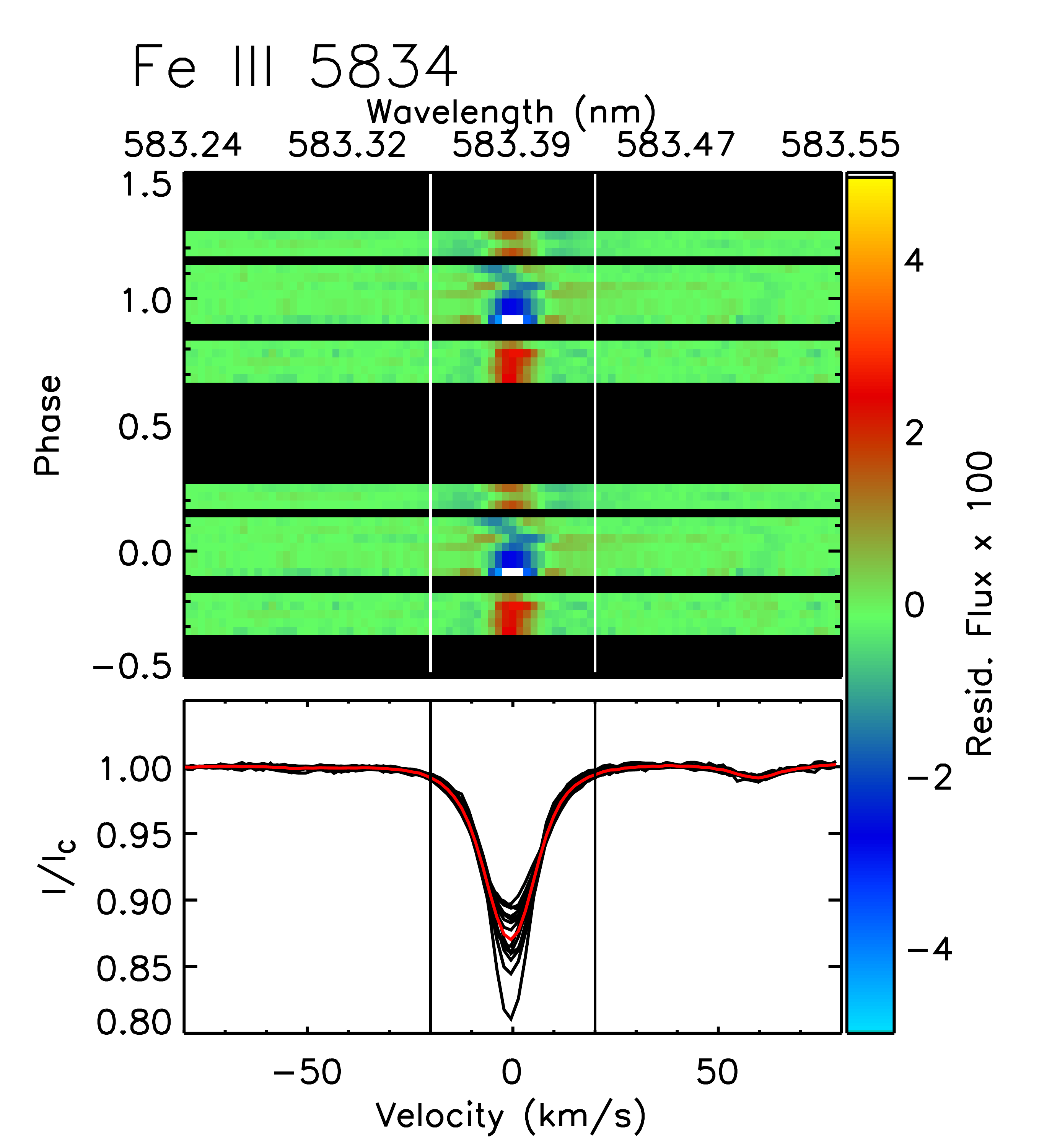} \\
\end{tabular}
    \caption{Dynamic spectra displaying line profile variations coherent with the rotational ephemeris of \citet{2017MNRAS.471.2286S,2018MNRAS.478L..39S}. Top panels show residual intensity mapped to colour as a function of rotational phase; bottom panels show phase-binned intensity spectra (black) and the mean reference spectrum (red).}\label{profiles}
\end{figure*}

To investigate this question, we calculated dynamic spectra for various spectral lines using the combined CORALIE and ESPaDOnS dataset. These are shown in Fig.\,\ref{profiles}. The data were first shifted to zero velocity by subtracting the measured radial velocity, phased assuming a 30-year rotational period with $T_0 = 2455219$ set by the time of maximum \bz, and then co-added in 30 phase bins. Radial velocity correction of individual spectra, combined with the presence of between 16 and about 100 spectra per bin, mean that pulsational variability should be removed to first order. For reference spectra we used the mean spectrum created from the full dataset. 

For reference, the top left panel of Fig.\,\ref{profiles} shows H$\alpha$; this line is already known to exhibit rotationally modulated magnetospheric emission \citep{2017MNRAS.471.2286S}. Maximum emission occurs at phase 0 \citep[i.e.\ at maximum \bz),][]{2017MNRAS.471.2286S,2018MNRAS.478L..39S} and the variability pattern is a smooth change in the strength of the central emission feature. 

Essentially all of the lines we examined display some form of rotationally modulated variation. The lines shown in Fig.\,\ref{profiles} were selected as exemplars of the three different patterns of variability. {He}\,{\sc ii} 4686\,\AA\ shows a very similar pattern of emission to H$\alpha$, suggesting that it is also partly formed within the magnetosphere. {O}\,{\sc ii} 4662\,\AA\ and {Si}\,{\sc iv} 4116\,\AA\  both show a pattern of alternating line strength between the cores and wings, with deeper line cores co-occuring with shallower line wings and vice-versa; the amplitude of the variation is furthermore similar between core and wings. {S}\,{\sc iii} 4362\,\AA\ and {Fe}\,{\sc iii} 5834\,\AA\ both show similar patterns of variation, but with the much more pronounced changes in line depth and relatively minor variation in the line wings. All of the metallic lines we investigated showed similar variations, with the lines at their deepest at phase 0, and at their shallowest near phases 0.25 and 0.75 (with the former definitely occurring near {\bz} $=0$).

It is not clear what the source of the metallic line profile variability is. Magnetospheric variation seems unlikely, since: 1) H$\alpha$ emission is very weak, implying negligible emission in metallic lines; 2) the metallic lines are at peak absorption at phase 0 when (if the variability were magnetospheric in origin) in-filling by emission should be at the greatest. Chemical spots also seem unlikely, as: 1) the formation of surface abundance inhomogeneities is inhibited in B0/B1 stars by their strong(er) winds; 2) all lines vary in essentially the same fashion, whereas the distribution of chemical spots tends to differ from one chemical element to another. Zeeman splitting may be plausible (the expected amplitude in a line with a Land\'e factor of 1.2, at 5000 \AA, for a star with a 1.2~kG surface magnetic field is about 2~\kms), however this does not seem to be the source of the variation: such splitting should be at its strongest (i.e.\ line width should be at a maximum) at maximum \bz, whereas precisely the opposite is the case. 

The difficulty in explaining $\xi^1$ CMa's rotationally modulated variation via conventional mechanisms suggests some heretofore unrecognized phenomenon. Whatever the origin of the rotationally modulated variation, however, in all cases it appears to be symmetrical about the line profile; thus, it should have no biasing affect upon the measurement of radial velocities.

\section{Conclusions}

We have analyzed spectroscopic and photometric data spanning over a century with the principal goal of examining the pulsation period evolution of the magnetic, slowly-rotating $\beta$~Cep star \xic. The observations confirm the previously-reported long-term increase of the pulsation period at a rate of approximately $0.3$\,s/cen, as well as recent, more rapid evolution corresponding to an approximate rate roughly three times larger. %\comment{Tony comments that the evidence for more rapid period evolution in the recent data is weak. GAW replies that this is incorrect. Review the preceding papers by Begy et al. and Shultz et al. to understand why.}

\xic\ exhibits a number of other characteristics that cause it to stand out from the broader population of $\beta$~Cep stars. The star exhibits a highly dominant  radial pulsation mode, and a century of observations reveals no clear evidence for change in pulsation amplitude. New TESS observations furthermore permit the detection of several low-amplitude modes with frequencies below 5~d$^{-1}$. As discussed in Sect.~\ref{RVcorr}, \xic\ exhibits a phase offset between maximum light and maximum RV of 0.334, significantly larger than the typical value of $\sim 0.25$. We demonstrate that these properties can be reconciled by a seismic model in which the star pulsates in the fundamental radial mode. It has the strongest magnetic field of any known $\beta$~Cep star, and it is one of the most slowly rotating known magnetic stars.

We conclude that the long-term lengthening of the period is not likely a consequence of a binary companion. That rate is however consistent with that expected from evolution of the star at its current  position on the main sequence inferred using standard stellar evolution models. We have no particular explanation for the recent, more rapid period evolution, although the associated timescale may be compatible with stellar rotation. Alternatively, we recall that we have observed only a very small part of a phenomenon which may take place on the nuclear timescale. Should we really expect it to proceed so smoothly? Given that the most recent observational data are more precise and provide a much denser temporal sampling, it may well be that similar short-term pulsational accelerations and decelerations have occurred in the past, and that they are a typical phenomenon. In fact, we know of no well-studied $\beta$~Cep star that shows period evolution with a constant $\dot P$ consistent with that expected from stellar evolution models. Period changes in BW~Vul, for example, were historically interpreted as a combination of $\dot P=$\,const and light-time effect, but the recent study by \citet{2012A&A...544A..28O} shows the reality can be (much) more complicated.  Continued high-cadence monitoring of the pulsation period of the star will be a key to understanding the roles and relationships of these properties in producing the observed period evolution.

\section*{Acknowledgments}
Based on data collected by the BRITE Constellation satellite mission, designed, built, launched, operated and supported by the Austrian Research Promotion Agency (FFG), the University of Vienna, the Technical University of Graz, the University of Innsbruck, the Canadian Space Agency (CSA), the University of Toronto Institute for Aerospace Studies (UTIAS), the Foundation for Polish Science \& Technology (FNiTP MNiSW), and National Science Centre (NCN). Based on observations obtained at the Canada-France-Hawaii Telescope (CFHT) which is operated by the National Research Council of Canada, the Institut national des sciences de l'Univers of the Centre national de la recherche scientifique of France, and the University of Hawaii. This paper includes data collected by the TESS mission. Funding for the TESS mission is provided by the NASA Explorer Program. Funding for the TESS Asteroseismic Science Operations Centre is provided by the Danish National Research Foundation (Grant agreement no.: DNRF106), ESA PRODEX (PEA 4000119301) and Stellar Astrophysics Centre (SAC) at Aarhus University. We thank the {\it TESS} team and staff and TASC/TASOC for their support of the present work. APi acknowledges support from the NCN grant 2016/21/B/ST9/01126, and helpful discussions with Przemek Walczak. Adam Popowicz was responsible for image processing and automation of photometric routines for the data registered by BRITE-nanosatellite constellation, and was supported by statutory activities grant SUT 02/010/BKM19 t.20. GAW acknowledges support from the Natural Sciences and Engineering Research Council (NSERC) of Canada in the form of a Discovery Grant. GH gratefully acknowledges funding through NCN grant 2015/18/A/ST9/00578. We thank Daniel Heynderickx for supplying his photometric data and Monika Rybicka for help with the APT data reduction. KZ acknowledges support by the Austrian Space Application Programme (ASAP) of the Austrian Research Promotion Agency (FFG). MES acknowledges support from the Annie Jump Cannon Fellowship, supported by the University of Delaware and endowed by the Mount Cuba Astronomical Observatory. The authors would like to thank Dr. V. Petit (University of Delaware) and the anonymous referee for helpful comments on the manuscript.

\bibliographystyle{mnras}
%\bibliography{article-v3} 
%\bibliography{article}
\bibliography{article}
%\bibliography{Xi1-CMa}
\bsp	% typesetting comment
\label{lastpage}
\end{document}